\documentclass[aps,prc,reprint,eqsecnum,showpacs,floatfix]{revtex4-2}

\usepackage{graphicx,epstopdf}
\usepackage{hyperref}
\usepackage{color}
\usepackage{subfigure}
\usepackage{amsmath,mathtools}
\usepackage{amssymb}
\usepackage{epsfig}
\usepackage{amsfonts}
\usepackage{bbm}
\usepackage{bm}
\usepackage{float}
\usepackage{xcolor}
\usepackage{bigints}

\def\bra#1{\bigl\langle{ #1} \bigr|}
\def\ket#1{\bigl|{ #1} \bigr\rangle}

\def\Ket#1{\Bigl|{ #1} \Bigr\rangle}
\def\ovlp#1#2{\bigl\langle{ #1}\big|{#2} \bigr\rangle}

\def\pd{{\phantom{\dagger}}} \def\rvec {{\bf r}} \def\pvec {{\bf p}}
\def\hvec {{\bf h}}
\def\qvec {{\bf q}}
\def\kvec {{\bf k}}

\def\creat#1{a_{#1}^{\dagger}}  
\def\annil#1{a_{#1}^{\pd}} 
 \def\Tr{{\cal T}\!r}
\def\LS{\widetildeto{{\bf L}\!\cdot\!{\bf S}}{{\bf L}\!\cdot\!{\bf S}}}
\def\LSp{\widetildeto{{\rm LS}}{{\rm LS}}'}

\def\LSsup{{\rm(LS)}}

\def\bsigma{{\bm\sigma}}
\def\balpha{{\bm\alpha}}
\def\tF{t_{\rm F}}
\def\1{\mathbbm{1}}

\def\KF{k_{\rm F}}
\def\KB{k_{\rm B}}
\def\SF{S_{\rm F}}
\def\S{{\bf S}}

\def\a0{a_0}
\def\I{{\rm i}}

\def\VLSq{{\tilde V}_{\rm p-h}^{\LSsup}}

 \def\Im{{\cal I}m}

\def\etal{{\em et al.\/}\ }
\def\ie{{\em i.e.\/}\ }
\def\cf{{\em cf.\/}\ }

\makeatletter
\def\mathcenterto#1#2{\mathclap{\phantom{#1}\mathclap{#2}}\phantom{#1}}
\let\old@widetilde\widetilde
\def\widetildeto#1#2{\mathcenterto{#2}{\old@widetilde{\mathcenterto{#1}{#2\,}}}}
\let\old@widehat\widehat
\def\widehatto#1#2{\mathcenterto{#2}{\old@widehat{\mathcenterto{#1}{#2\,}}}}
\makeatother

\begin{document}

\title{Variational and parquet-diagram calculations for neutron
  matter.\\ V. Triplet pairing}

\author{E.~Krotscheck$^{\dagger\ddagger}$, P. Papakonstantinou$^+$, and J. Wang$^\dagger$}

\affiliation{$^\dagger$Department of Physics, University at Buffalo,
  SUNY Buffalo NY 14260} \affiliation{$^+$Rare Isotope Science
  Project, Institute for Basic Science, Daejeon 34000, Korea}
  \affiliation{$^\ddagger$Institut f\"ur Theoretische Physik, Johannes
    Kepler Universit\"at, A 4040 Linz, Austria}

\begin{abstract}

We apply a large-scale summation of Feynman diagrams, including the
class of parquet-diagrams {\em plus\/} important contributions outside
the parquet class, for calculating effective pairing interactions and
subsequently the superfluid gap in P-wave pairing in neutron matter.
We employ realistic nucleon-nucleon interactions of the $v_8$ type and
perform calculations up to a Fermi momentum of $1.8\,$fm$^{-1}$.  We
find that many-body correlations lead to a strong reduction of the
spin-orbit interaction, and, therefore, to an almost complete
suppression of the $^3$P$_2$ and $^3$P$_2$-$^3$F$_2$ gaps. We also
find pairing in $^3$P$_0$ states; the strength of the pairing gap
depends sensitively on the potential model employed.  Our results for
triplet pairing are relevant for assessing superfluidity in
neutron star interiors, whose presence can affect the cooling of
neutron stars.

\end{abstract}
\maketitle
\section{Introduction}

Pairing is manifest in the spectra of finite nuclei, as recognized
long ago~\cite{BMP}, and is predicted to develop at various densities
in nucleonic matter found in compact
stars~\cite{DeH2003,PageReddy2006,BCS50Book,Gezerlis2014,StellarSuperfluids}.
The most basic type of calculations of pairing and superfluidity or
superconductivity in nuclear matter rely on mean-field-like (or
Hartree-Fock like) implementations of Bardeen-Cooper-Schrieffer (BCS)
theory.  It was recognized long ago that the BCS equations can be
solved not only for soft interactions, but also for interactions with
a repulsive hard core~\cite{CMS}.  To what extent a mean-field theory
captures the essential physics of the nucleonic system is an important
question.

In the case of pure neutron matter, theoretical approaches generally
agree that the value of the S-wave pairing gap value should reach a
maximum of roughly 2~MeV (within about 1~MeV) at a Fermi momentum
somewhat below 1~fm$^{-1}$, corresponding to subsaturation
densities~\cite{Wam93,Gezerlis2014,StellarSuperfluids,50yrs,%
  Pavlou2017,JLTP189_234,JLTP189_250,Schwenk03,Fabrocinipairing,Cao2006,Drischler2017,v3bcs}.
There still remain some quantitative discrepancies. This is not
surprising given the extreme sensitivity of the pairing gap on the
interaction and also in view of the different ways effects beyond
mean-field theory are taken into account, if at all. Similar
conclusions seem valid for $\beta$-equilibrated matter
\cite{PhysRevC.103.025807}.

The situation in the case of triplet P-wave pairing in neutron matter
is much more uncertain. It is not even clear whether a pairing gap
develops in this channel.
The question is of particular astrophysical significance since triplet
pairing is expected to be most favored at densities found in the outer
liquid core of neutron stars and its presence would affect the cooling
curves of neutron stars, {\em i.e.}, the evolution of surface
temperature with time~\cite{PageReddy2006,StellarSuperfluids,YaH2003}.
Clarifying the question of triplet pairing is necessary in order to
make the most of modern observational capabilities.

Microscopic calculations employing realistic nuclear potentials
generally predict, at the mean field level, a non-vanishing triplet
pairing
gap~\cite{TaT1993,EEH1996,JLTP189_234,JLTP189_250,Drischler2017}.
However, at the high densities involved, effects beyond mean-field
approximations cannot be neglected.  For example, it has been found
that realistic nucleon effective masses~\cite{TaT1993} and short-range
correlations~\cite{PhysRevC.94.025802,apjlett} can reduce or eliminate
the gap \cite{JLTP189_234}. Three-nucleon interactions of the Urbana IX family
can also lead to vanishing gaps~\cite{Yuan2007,JLTP189_361}. The
  effect of three-nucleon interactions derived within chiral effective
  field theory ($\chi$EFT) has been studied, too, and found
  regulator-dependent but potentially significant in the triplet
  channel \cite{JLTP189_234,Drischler2017}.

Studies of polarization and screening effects in the triplet channel
have been scarce.  The in-medium spin-orbit and tensor interaction
components are especially important, but they are not well understood.
Even working with bare nucleon-nucleon potentials, the results for the
triplet channel depend on the interaction type ({\em e.g.,} Argonne,
Bonn, $\chi$EFT, {\em etc}).  A study of polarization contributions to
the spin-dependent nuclear interaction in the medium has suggested a
suppression of the $^3$P$_2$ gap~\cite{SchwenkFriman2004}.

It is the purpose of this work to apply a comprehensive diagrammatic
theory to the problem of triplet pairing in neutron matter.  Our
method is based on an evaluation of Feynman diagrams that include
self-consistently ring and ladder diagrams, {\em i.e.}, the parquet
class, but also totally irreducible diagrams, dubbed ``twisted
chains'' which become important when the interactions between
particles in spin-singlet and spin-triplet states are very
different. The method has been developed and applications were
presented in a series of previous
papers~\cite{v3eos,v3twist,v3bcs,v4}.  In Ref.~\onlinecite{v3bcs}, it
was applied to the pairing gap problem in the singlet channel and the
effect of the correlations was discussed. The method was generalized
in Ref. \onlinecite{v4} to spin-orbit forces. It was demonstrated that
many-body correlations affect dramatically the in-medium spin-orbit
component, which in turn affects the triplet pairing gap.  The
above result forms the motivation for the present work.

This paper is organized as follows.  The theoretical background and
details of the formalism for $v_8$ bare potentials are presented in
Sec.~\ref{sec:methods}.  Results are presented and discussed in
Sec.~\ref{sec:results}.  We summarize our findings in
Sec.~\ref{sec:summary}.  The Appendices contain additional information
and technical details as well as some results with bare interactions
for comparison with earlier work.

\section{Microscopic Theory}
\label{sec:methods}
\subsection{Interactions: Semi-empirical nucleon-nucleon forces in operator representation}
\label{ssec:interactions}

Accurate representations of the nucleon-nucleon potentials
\cite{Reid68,Bethe74,AV18,Reid93} are constructed to fit the
interactions in each partial wave to scattering data and deuteron
binding energies. For the purpose of identifying specific physical
effects and of high-level many-body calculations, an interaction given
in the form of a sum of local functions times operators acting on the
spin, isospin, and possibly the relative angular momentum variables of
the individual particles is preferred \cite{Day81,AV18,Wiri84}, \ie,
\begin{equation}
\hat v (i,j) = \sum_{\alpha=1}^n v_\alpha(r_{ij})\,
        \hat O_\alpha(i,j),
\label{eq:vop}
\end{equation}
where $r_{ij}=\left|\rvec_i-\rvec_j\right|$ is the distance between
particles $i$ and $j$. According to the number of operators $n$, the
potential model is referred to as a $v_n$ model
potential. Semi-realistic models for nuclear matter keep at least the
six to eight base operators
\begin{eqnarray}
\hat O_1(i,j;\hat\rvec_{ij})
        &\equiv& \hat O_{\rm c} = \1\,,
\nonumber\\
\hat O_3(i,j;\hat\rvec_{ij})
        &\equiv& {\bsigma}_i \cdot {\bsigma}_j\,,
\nonumber\\
\hat O_5(i,j;\hat\rvec_{ij})
&\equiv& S_{ij}(\hat\rvec_{ij})\nonumber\\
      &\equiv& 3({\bm\sigma}_i\cdot \hat\rvec_{ij})
      ({\bsigma}_j\cdot \hat\rvec_{ij})-{\bsigma}_i \cdot {\bsigma}_j\,,
      \nonumber\\
  \hat O_7(i,j;\rvec_{ij},\pvec_{ij})
  &\equiv&\rvec_{ij}\times\pvec_{ij}\cdot\S\,,
  \nonumber\\
      \hat O_{2\alpha}(i,j;\hat\rvec_{ij}) &=& \hat O_{2\alpha-1}(i,j;\hat\rvec_{ij})
      {\bm\tau}_i\cdot{\bm\tau}_j\,,
  \label{eq:operator_v8}
\end{eqnarray}
where $\S\equiv\frac{1}{2}(\bsigma_i+\bsigma_j)$ is the total spin,
and $\pvec_{ij}=\frac{1}{2}(\pvec_i-\pvec_j)$ is the relative momentum
operator of the pair of particles. In the following, we will also use the notation 
$\alpha \in \{({\rm cc}), (\rm{c}\tau), (\sigma{\rm c}),
(\sigma\tau), ({\rm Sc}),  ({\rm S}\tau) , ({\rm LSc}) , ({\rm LS}\tau) \}$
for $\alpha=1-8$.
In neutron matter, the operators
are projected to the isospin=1 channel, \ie we have
\begin{equation}
  O_\alpha(i,j,\hat\rvec_{ij}) \rightarrow  O_\alpha(i,j,\hat\rvec_{ij})+O_{\alpha+1}(i,j,\hat\rvec_{ij})
\end{equation}
for odd $\alpha$ and $O_\alpha(i,j,\hat\rvec_{ij})=0$ for even $\alpha$.
The new set of interaction channels will be $\alpha \in \{({\rm c}), (\sigma),
 ({\rm S}), ({\rm LS})\}$.

The Argonne interaction \cite{AV18} is, among others, formulated in
the operator representation. For the $v_8$ version of the Reid 68
interaction we have taken the six components $V^{(6)}_{\alpha}(r)$,
$\alpha \in \{({\rm cc}), (\rm{c}\tau), (\sigma{\rm c}),
(\sigma\tau), ({\rm Sc}),  ({\rm S}\tau) \}$ from Eqs. (A3)-(A8) of Ref. \onlinecite{Day81}. The
spin-orbit components $\alpha \in \{  ({\rm LSc}) , ({\rm LS}\tau) \}$ 
were constructed \cite{v4}, following the procedure of
Ref. \onlinecite{AV18}, from the isospin $T=0$ and isospin $T=1$ components of
the Reid interaction, \cf Eqs. (20) and (30) of
Ref. \onlinecite{Reid68}.

A somewhat more recent interaction \cite{Reid93} is given only in a
partial wave representation.  To derive an operator structure of the
form \eqref{eq:vop} from the partial wave representation of the
interaction we  need the partial-wave representation of the
operators ${\bf L}\cdot{\bf S}$ and $\S_{12}$. For neutron
matter we only need the case $T=1$. 
On the other hand, we need a definite total spin S=0 or S=1 for the pairing problem.
The central components of the interaction are then
\begin{subequations}
\begin{eqnarray}
  v_{\rm c,0}(r) &=& v_c(r) - 3 v_\sigma(r)\,,\\
  v_{\rm c,1}(r) &=& v_c(r) +  v_\sigma(r)\,.
\end{eqnarray}
\end{subequations}

In the spin-singlet case we only
have a central interaction and the only reasonable choice is
\begin{equation}
V(S=0,T=1)(r) \equiv V_{^1S_0}(r) = v_{\rm c,0}(r)
\end{equation}

In the spin-triplet case, the operator structure
is 
\begin{equation}
\hat V(S=1,T=1)(r) = v_{\rm c,1}(r) \1 + v_{\rm S}(r)\hat S_{12}(\rvec)
+ v_{\rm LS}(r) {\bf L}\cdot{\bf S}
\end{equation}
where, for each partial wave, we have \cite{Naghdi2014}
\begin{equation}
{\bf L}\cdot{\bf S}=\begin{pmatrix}
j-1 & 0 \\
0 & -j-2 \end{pmatrix}\,.\label{eq:LS}
\end{equation}
and
\begin{equation}
S_{ij} = \begin{pmatrix}
\frac{-2(j-1)}{2j+1} & \frac{6\sqrt{j(j+1)}}{2j+1}\\
\frac{6\sqrt{j(j+1)}}{2j+1} & \frac{-2(j+2)}{2j+1}\end{pmatrix}\,.\label{eq:Sij}
\end{equation}

It is evidently impossible to represent all individual partial waves
with only eight (or, in neutron matter, four) interaction components
as required by the operator representation \eqref{eq:vop}. We have
therefore explored two possibilities, as follows, to define an operator
representation of the Reid93 potential. The tensor interaction is in
both cases determined by the $^3{\rm P}_2- ^3{\rm F}_2$ interaction,
\begin{equation}
  v_{\rm S}(r) = \frac{5}{6\sqrt{6}} V_{^3{\rm P}_2- ^3{\rm F}_2}(r)\,.
  \label{eq:Vtensor}
  \end{equation}

\begin{itemize} 
\item 
If we want to reproduce the $^3$P$_0$-$^3$P$_0$ and the
$^3$P$_2$-$^3$P$_2$ phase shifts, we get
\begin{subequations}\label{eq:R93a}
  \begin{eqnarray}
  v_{\rm c,1}(r) &=& \frac{1}{3}\left(2V_{^3{\rm P}_2- ^3{\rm P}_2}(r)+
  V_{^3{\rm P}_0- ^3{\rm P}_0}(r)\right)+\frac{8}{5}v_{\rm S}(r)\nonumber\\
  \\
  v_{\rm LS}(r) &=& \frac{1}{3}\left(V_{^3{\rm P}_2- ^3{\rm P}_2}(r)-
  V_{^3{\rm P}_0- ^3{\rm P}_0}(r)\right)-\frac{6}{5}v_{\rm S}(r)\nonumber\\
  \end{eqnarray}
\end{subequations}
\item 
If we want to reproduce the  $^3$P$_2$-$^3$P$_2$ and the
$^3$F$_2$-$^3$F$_2$ phase shifts, we have
\begin{subequations}\label{eq:R93b}
  \begin{eqnarray}
  v_{\rm c,1}(r) &=& \frac{1}{5}\left(4V_{^3{\rm P}_2- ^3{\rm P}_2}(r)+
  V_{^3{\rm F}_2- ^3{\rm F}_2}(r)\right)+\frac{16}{25}v_{\rm S}(r)\nonumber\\
  \\
  v_{\rm LS}(r) &=& \frac{1}{5}\left(V_{^3{\rm P}_2- ^3{\rm P}_2}(r)-
  V_{^3{\rm F}_2- ^3{\rm F}_2}(r)\right)-\frac{6}{25}v_{\rm S}(r).\nonumber\\
  \end{eqnarray}
\end{subequations}
\end{itemize} 
We shall refer to the approximations \eqref{eq:R93a} and
\eqref{eq:R93b}\ as ``Version a'' and ``Version b'',
respectively. Turning the ambiguity into an advantage, we shall
compare the results from these two interaction models to assess the
accuracy of predictions based on the operator representation of the
Reid 93 interactions.  For the body of this work we have chosen the
operator form \eqref{eq:R93a} that reproduces the lowest-lying partial
waves.  The equation of state obtained with this interaction is
practically indistinguishable from the equation of state obtained from the
Reid 68 and the Argonne $v_8$ potential. Nevertheless, exploring
different representations can provide insight into the robustness of
theoretical predictions, see Appendix \ref{app:baregaps}.

We have focused in this paper on calculations for 2 variants of the
Reid interaction \cite{Reid68,Reid93} as well on the $v_8$ truncation
of the Argonne interaction \cite{AV18}. This was done to be consistent
with our previous work \cite{v3eos,v3twist,v3bcs} as well as other
microscopic calculations
\cite{JWCgap,KKC96,KhodelClark2001,Baldo2012}.  We are, of course,
aware of the fact that there are more modern interactions
\cite{RevModPhys.81.1773,Machleidt} which have been used for the
calculation of properties of neutron and nuclear matter
\cite{PhysRevLett.111.032501,PhysRevLett.113.192501,PhysRevLett.120.052503,PhysRevC.91.024003}.
The present paper focuses on P-wave paring, whereas an investigation
of these new interactions warrants a much broader investigation
including other phenomena such as the structure \cite{v3eos}, S-wave
pairing \cite{v3bcs}, the dynamic structure \cite{v4}, and the optical
potential. We shall address these issues in future work.

  \subsection{Jastrow-Feenberg variational method and parquet-diagrams}
\label{ssec:JFandparquet}

In terms of the paradigms of perturbative many-body theory
\cite{BaymKad}, it is easy to argue that the {\em minimum\/} set of
Feynman diagrams for a trustworthy microscopic treatment of strongly
interacting systems is the set of {\em parquet-diagrams\/}
\cite{parquet1,JWE93}. While the insight into what is needed is quite
obvious, the execution of such a program is far from trivial. One must
seek approximations, but such steps are ambiguous without
further guidance.

An approach that is superficially very different from perturbative
many-body theory has been suggested by Jastrow \cite{Jastrow55} and
Feenberg \cite{FeenbergBook}.  For simple, state-independent
interactions as appropriate for electrons or quantum fluids, the
Jastrow-Feenberg ansatz \cite{Jastrow55,FeenbergBook} for the wave
function
\begin{equation}
\Psi_0 = F_N\Phi_0,\qquad F_N = \prod^N_{\genfrac{}{}{0pt}{1}{i,j=1}{i<j}}  f(r_{ij})
\label{eq:Jastrow}
\end{equation}
and its logical generalization to multi-particle correlation functions
has been extremely successful. Here, $\Phi_0$ is a model state
describing the statistics and, when appropriate, the geometry of the
system. For fermions, it is normally taken as a Slater determinant. We
will here use the generalization to Bardeen-Cooper-Schrieffer (BCS)
states
\cite{YangClarkBCS,HNCBCS,Fantonipairing,Fabrocinipairing,Fabrocinipairing2,fullbcs,v3bcs}.

One of the reasons for the success of this method is that it
provides an upper bound for the ground state energy
\begin{equation}
E_0 = \frac{\bra{\Psi_0}H\ket{\Psi_0}}{\ovlp{\Psi_0}{\Psi_0}}\,.
  \label{eq:energy}
\end{equation}
A singularly useful hierarchy of equations for the calculation of the
energy expectation value \eqref{eq:energy} is the hypernetted-chain
summation technique \cite{Morita58,LGB}; it is characterized by the
fact that it allows, at every level of implementation, the unconstrained
optimization of the correlations via the variational principle
\begin{equation}
\frac{\delta E_0}{\delta f}({\bf r}_i,{\bf r}_j) = 0\,.
\label{eq:euler}
\end{equation}
The method is referred to as the
(Fermi-)Hypernetted-Chain-Euler-Lagrange, (F)HNC-EL, procedure.

An important insight was that this procedure corresponds to a
summation of a ``local approximation'' of the parquet diagrams
\cite{parquet1,parquet2,parquet3}. This was proven first for bosons;
some additional approximations are made in a Fermi system
\cite{fullbcs}.  The variational problem \eqref{eq:euler} ensures that
one uses the best approximation for the computational effort one is
willing to spend. We shall therefore use the language of
Jastrow-Feenberg and parquet-diagrams interchangeably, in particular
for the benefit of those readers who are less familiar with the
former.

The situation is considerably more complicated for realistic
nucleon-nucleon interactions of the form (\ref{eq:vop}). A plausible
generalization of the wave function (\ref{eq:Jastrow}) is the
``symmetrized operator product (SOP)'' \cite{FantoniSpins,IndianSpins}
\begin{equation}
        \Psi_0^{{\rm SOP}} = F_N^{{\rm SOP}}  \Phi_0\,,\qquad
        F_N^{{\rm SOP}} = {\cal S}
        \Bigl[ \prod^N_{\genfrac{}{}{0pt}{1}{i,j=1}{i<j}} \hat f (i,j)\Bigr]
\label{eq:f_prodwave}
\end{equation}
where
\begin{equation}
  \hat f(i,j) = \sum_{\alpha=1}^n f_\alpha(r_{ij})\,
  \hat O_\alpha(i,j)\,,
  \label{eq:fop}
\end{equation}
and ${\cal S}$ stands for symmetrization. The symmetrization is
necessary because the operators $\hat O_\alpha(i,j)$ and $\hat
O_\beta(i,k)$ do not necessarily commute. We have highlighted recently
\cite{v3twist} (see also Ref. \onlinecite{SpinTwist}) the importance
of a proper symmetrization in cases where the bare interaction is
different in spin-singlet and spin-triplet channels; we shall return
to this point in section \ref{sssec:twists}.

In a preceding series of papers \cite{v3eos,v3twist,v4}, we have
developed practical and efficient methods for the summation of the
parquet diagrams, including the most important commutator diagrams
mentioned above.  We shall briefly review the resulting equations in
the next sections.

\subsubsection{Diagram summation: Chain diagrams}
\label{sssec:chains}

One of the components of parquet-diagram theory is the summation of
the chain diagrams.  We assume a {\em local\/} effective particle-hole
interaction of the same form as the bare interaction \eqref{eq:vop}
which is given, in momentum space
\begin{equation}
  \hat V_{\rm p-h}(q) = \sum_{\alpha=1}^8 \tilde V^{(\alpha)}_{\rm p-h}(q)\,
        \hat O_\alpha(i,j)\,.
\label{eq:vphop}
\end{equation}
$\hat V_{\rm p-h}(q)$ is, in the long wavelength limit, related to
Landau's Fermi-Liquid interactions
\cite{LandauFLP1,LandauFLP2,BaymPethick} or, at finite wave numbers,
to pseudopotentials \cite{ALP78,Wam93}. The momentum-space components of the
interactions in the different operator channels are the Fourier transforms
\begin{equation}
  \tilde V_{\rm p-h}^{(\alpha)}(q) =\begin{cases}
  \phantom{-}\rho \int d^3r V_{\rm p-h}^{(\alpha)}(r)
  j_0(qr)&\ \mathrm{for}\ \alpha = 1\ldots 4,\\
   -\rho \int d^3r V_{\rm p-h}^{(\alpha)}
   (r)j_2(qr)&\ \mathrm{for}\ \alpha = 5,6\,.\\
   \phantom{-}\frac{\rho}{2}\int d^3r V_{\rm
    p-h}^{\alpha}(r) r\KF j_1(qr)&\ \mathrm{for}\ \alpha = 7,8\,,
   \end{cases}
\end{equation}
where the momentum representation of the tensor operator is obtained
by replacing $\hat r_{ij}\rightarrow\hat \qvec$.  As usual, we have
above defined the Fourier transforms with a density factor such that
the momentum space interactions also have the dimension of energy.

The summation of chain diagrams is best carried out by transforming
the spin and tensor operators into the longitudinal and transverse
operators \cite{FrimanNyman,WEISE1977402}. The particle-hole interaction
is then a linear combination of the four operators
\begin{subequations}
  \label{eq:Qdef}
  \begin{eqnarray}
    \hat  Q_1&\equiv&\1\,\label{eq:Q1} \, \\
  \hat Q_3 &\equiv& \hat L(\hat\qvec) = ({\bsigma}\cdot \hat\qvec)({\bsigma}'\cdot \hat\qvec)\,,\\
  \hat Q_5 &\equiv& \hat T(\hat\qvec) = {\bsigma}\cdot{\bsigma'}-({\bsigma}\cdot \hat\qvec)({\bsigma}'\cdot \hat\qvec)\,,\\
  \hat Q_7 &\equiv& \LS \equiv
  \frac{\I}{\KF}\hat
  \qvec\times\Delta\hvec\cdot\S\label{eq:LSdef}\,.
\end{eqnarray}
\end{subequations}
The operator $\LS$ acts
only in spin-space. It depends parametrically on the direction
$\hat\qvec$ of momentum transfer and the difference of the hole wave
numbers $\Delta \hvec\equiv \hvec-\hvec'$. We shall generally mean the
{\em momentum space\/} representations \eqref{eq:Qdef}, when we refer
to the operators $\hat Q_\alpha$.

A complete derivation of chain-diagram summations including the
spin-orbit interaction was done in Ref. \onlinecite{v4}.  The sum of
all chain diagrams can no longer be represented as a linear
combination of momentum-space functions times the operators
\eqref{eq:Qdef}. Three more operators are needed. The coefficient
functions of those additional operators that contain an even number of
spin-orbit operators are, however, of order $\left[\VLSq(q)\right]^2$
and are numerically very small. An interesting feature of the
spin-orbit order term, which might be relevant in different physical
circumstances, will be outlined below in connection with
Eqs. \eqref{eq:VLSlocal}-\eqref{eq:VLSp}.

To represent the sum of all chain diagrams, it has turned out convenient to
introduce
\begin{subequations}
  \begin{eqnarray}
    \tilde V_{\rm p-h}^{\rm (c)}(q;\omega)&\equiv& \tilde V_{\rm p-h}^{\rm (c)}(q)+
    \frac{1}{4}\chi_0^{(\perp)}(q;\omega)\left[\VLSq(q)\right]^2\!\!,\nonumber\\
    &&\label{eq:Vcredef}\\
    \tilde V_{\rm p-h}^{\rm (T)}(q;\omega)&\equiv&
    \tilde V_{\rm p-h}^{\rm (T)}(q)+
    \frac{1}{8}\chi_0^{(\perp)}(q;\omega)\left[\VLSq(q)\right]^2\!\!,\nonumber\\
    &&\label{eq:VTredef}\\
    \tilde V_{\rm
      p-h}^{\rm (L)}(q;\omega) &\equiv& \tilde V_{\rm p-h}^{\rm (L)}(q)\,.
    \label{eq:VLredef}
  \end{eqnarray}
\end{subequations}
We have defined above a 
transverse Lindhard function
\begin{equation}
  \chi_0^{(\perp)}(q;\omega)=\frac{1}{N}\Tr_\bsigma\sum_\hvec
  \left|\frac{\hat \qvec\times\hvec}{\KF}\right|^2\frac{2(\varepsilon_p-\varepsilon_h)}
  {(\hbar\omega+\I\eta)^2 - (\varepsilon_p-\varepsilon_h)^2}\label{eq:chi0trans}
\end{equation}
where $\qvec=\pvec-\hvec$. The $\varepsilon_p$ and $\varepsilon_h$ are
the single particle energies of correlated basis functions (CBF)
theory that have been discussed elsewhere
\cite{Johnreview,polish,KroTrieste}, see also the Appendix of
Ref. \onlinecite{v3bcs}.

The sum of all chain diagrams containing an odd number of spin-orbit
operators can be written as
\begin{subequations}
\begin{align}
\hat W_{\rm LS}^{(\rm odd)}(\qvec;\omega)&=W^{\rm (LS)}(\qvec,\omega) \LS +
W^{(\rm LS')}(\qvec,\omega) \LSp\\
W^{(\rm LS)}(\qvec,\omega)  &= \frac{1}{2}
  \frac{\VLSq(q)}{1-\chi_0(q;\omega)\tilde V_{\rm p-h}^{\rm (c)}(q;\omega)}\nonumber\\
  &+ \frac{1}{2}\frac{\VLSq(q)}{1-\chi_0(q;\omega)\tilde V_{\rm p-h}^{\rm (T)}(q;\omega)}\label{eq:VLSlocal}\\
  W^{(\rm LS')}(\qvec,\omega) &=
  \frac{1}{2}
  \frac{\VLSq(q)}{1-\chi_0(q;\omega)\tilde V_{\rm p-h}^{\rm (T)}(q;\omega)}\nonumber\\
  &- \frac{1}{2}\frac{\VLSq(q)}{1-\chi_0(q;\omega)\tilde V_{\rm p-h}^{\rm (c)}(q;\omega)}
 \label{eq:VLSodd}\,,
\end{align}
where $\chi_0(q;\omega)$ is the usual Lindhard function, and we deviate
here slightly from the definition in Ref. \onlinecite{v4}
\begin{equation}
\LSp =  \frac{\I}{2\KF}\hat\qvec\times(\hvec+\hvec')
\cdot(\bsigma-\bsigma')\label{eq:VLSp} \,.
\end{equation}
\end{subequations}
The operator $\LSp$ is antisymmetric in the spins and does not
contribute to the pairing interactions. In different physical
circumstances the term proportional to $\LSp$ could be very
interesting because it is the only term in the effective interaction
that couples spin-singlet and spin-triplet states.

The energy dependent effective interaction can then be represented
by the operator expansion
\begin{equation}
  \hat W(\qvec;\omega) =
  \sum_{\substack{\alpha\,{\rm odd}}}^7 \tilde W^{(\alpha)}(q;\omega)\hat Q_\alpha \,,
\end{equation}
where
\begin{eqnarray}
  \tilde W^{(\alpha)}(q;\omega)
  &=& \frac{\tilde V_{\rm p-h}^{(\alpha)}(q;\omega)}
  {1-\chi_0(q;\omega)\tilde V_{\rm p-h}^{(\alpha)}(q;\omega)}\quad
  \mathrm{for}\quad \alpha = 1, 3, 5,\nonumber\\
  &&\label{eq:WindRPA}
\end{eqnarray}
and Eq. \eqref{eq:VLSlocal} for $\alpha=7$.

\subsubsection{Diagram summation: Ladder diagrams}
\label{sssec:ladders}

The second component of the parquet (or (F)HNC-EL) summation is the
summation of ladder diagrams. That is generally accomplished by the
Bethe-Goldstone equation. In its most primitive form the equation
contains, as the only many-body effect, the Pauli exclusion principle.
Summing parquet diagrams amounts to supplementing the bare interaction
$v(r)$ by an ``induced interaction'' $\tilde w_I(q,\omega)$.  This induced
interaction is energy dependent. {\em Local\/} parquet theory then
replaces this interaction by an energy-independent interaction which
is defined such that the static approximation leads to the same
(observable) static structure function as the dynamic interaction:
\begin{eqnarray}
    S(q)
    &=& -\int_0^\infty \frac{d\hbar\omega}{\pi} \Im
 \frac{\chi_0(q;\omega)} {1-\tilde V_{\rm
     p-h}(q)\chi_0(q;\omega)}\nonumber\\
 &=& -\int_0^\infty \frac{d\hbar\omega}{\pi} \Im
 \left[\chi_0(q;\omega)+ \chi^2_0(q;\omega) \widetilde W(q;\omega)\right]
 \nonumber\\
  &\overset{!}{=}& 
-\int_0^\infty \frac{d\hbar\omega}{\pi} \Im
\left[\chi_0(q,\omega)+ \chi^2_0(q,\omega) \widetilde W(q)\right]\,,
\nonumber\\
\label{eq:Wlocal}
\end{eqnarray}
which defines the energy dependent  induced interaction though
\begin{equation}
\widetilde
W(q,\omega) = \tilde V_{\rm p-h}(q) + \tilde w_I(q,\omega)\label{eq:wIomega}
\end{equation}
and its {\rm local approximation
  \begin{equation}
\widetilde
W(q) = \tilde V_{\rm p-h}(q) + \tilde w_I(q)
\label{eq:wIdef}\,.
\end{equation}

The Bethe-Goldstone equation defines a pair wave function
$\psi(\rvec)$ which can be identified, in the simplest case of central
state-independent correlations and the FHNC//0 or parquet//0
approximation, with the ``direct correlation function'' $\Gamma_{\rm
  dd}(r)$ through
\begin{equation}
  1 + \Gamma_{\rm dd}(\rvec) = |\psi(\rvec)|^2
\end{equation}
and is, in that approximation, related to the static structure
function through
\begin{equation}
  \widetilde \Gamma_{\rm dd}(\qvec) = \frac{\SF(q)-S(q)}{\SF^2(q)}\,.
  \end{equation}
Here, $\SF(q)$ is the static structure function of non-interacting fermions.
For the present case of interacting nucleons, all quantities
are operators in the same basis
as the microscopic interaction,
see Refs. \onlinecite{v3eos} and \onlinecite{SmithSpin} for details.
The resulting particle-hole interaction is
\begin{eqnarray}
  \hat V_{\rm p-h}(\rvec) &=& \frac{\hbar^2}{m}\left|\nabla \hat \psi(\rvec)\right|^2
  \\
  &+&\hat \psi^*(\rvec) [\hat v(\rvec) + \hat V_I(\rvec) + \hat w_I(\rvec)]
    \hat \psi(\rvec)
      - \hat w_I(r)\,.\nonumber
\end{eqnarray}
The additional term $\hat V_I(r)$ is the ``irreducible'' interaction
which arises from the beyond-parquet contributions.

\subsubsection{Diagram summation: Beyond parquet}
\label{sssec:twists}

The most notorious problem of an operator-dependent variational wave
function \eqref{eq:f_prodwave} is that symmetrization must be carried
out explicitly in order to get a valid variational principle
\eqref{eq:euler}.  Light was shed on the meaning of the arising
``commutator contributions'' by Smith and Jackson \cite{SmithSpin} who
showed, for a fictitious system of bosons with spin, isospin, and
tensor forces, that the parquet-diagram summation leads to an optimized
Bose-version of Ref. \onlinecite{FantoniSpins}, \ie to a theory where
{\em all commutator diagrams are omitted.\/} The conclusion is
therefore that the wave function \eqref{eq:f_prodwave} contains {\em
  more\/} than just the parquet class; non-parquet diagrams simply are
neglected when commutators are neglected.

From the point of view of the variational wave function
\eqref{eq:f_prodwave}, it is abundantly clear that commutator diagrams
are important whenever the interaction in singlet and triplet channels
is very different: Working out the simplest non-trivial commutators
leads to contributions to the energy where the spin-singlet
interaction is multiplied by a spin-triplet correlation function
$f_{\rm triplet}(r)$ and vice versa. Taking the extreme case of
hard-core interactions with different core sizes for singlet and
triplet states \cite{Ohmura56}, and simplistic correlations functions
as they are being used in low-order constraint variational (LOCV)
calculations, would lead to divergences.

The issue is less obvious in diagrammatic perturbation theory. But
once the importance of non-parquet diagrams is realized, diagrammatic
perturbation theory offers an intuitive explanation.  We show in
Fig. \ref{fig:ladders} the simplest possibility of ``twisting'' chain
diagrams.  The leftmost diagram is the ordinary second-order ladder
diagram as summed by the Bethe-Goldstone equation. The middle diagram
shows the simplest case where the bare interaction is replaced by the
induced interaction $\tilde w_I(q,\omega)$ or, in practice, its local approximation
$\tilde w_I(q)$. In both cases, if a pair of
particles enters with quantum numbers $\ket{\kvec_1,\kvec_1',S}$ {\em
  then it remains in that spin configuration throughout the process.}

The third diagram, although it has the same
components, is by definition not a parquet diagram; it represents the
simplest contribution to the irreducible interaction $\hat V_I(r)$.
Working out the spin-flux one obtains for that
\begin{eqnarray}
  \left(\Delta V_I\right)^{(2)}(k) &=& \int \frac{d^3q}{(2\pi)^3\rho}
  \frac{1}{E(k,q)}\biggl[\tilde w_{I,c}(q)\hat V(\kvec-\qvec)
  \nonumber\\
  &+& \tilde w_{I,\sigma} (q)\left[2\tilde V_t(\kvec-\qvec)-
  \tilde V_s(\kvec-\qvec)\right]P_t\nonumber\\
  &-& 3\tilde w_{I,\sigma} (q)\tilde V_t(\kvec-\qvec)P_s\biggr]\,,
  \label{eq:vI2}
\end{eqnarray}
where $E(\kvec,\qvec)$ is the energy denominator appropriate for the
process, and we have extended second order chains to the fill induced
interaction. Also, we have here omitted the tensor- and spin-orbit
force for brevity.

The first line in the expression \eqref{eq:vI2} simply says that, if
the induced interaction does not carry spin, then the spins of the
interaction operator $\hat V(\kvec-\qvec)$ remain the same.
The second line of \eqref{eq:vI2} carries our message:
If the induced interaction carries a spin, then the spin-triplet
interaction will contribute to the spin-singlet component of the
induced interaction. Similarly, the spin-singlet interaction contributes to
the spin-triplet component of the
induced interaction.

\begin{figure}
  \centering
    \includegraphics[width=0.95\columnwidth]{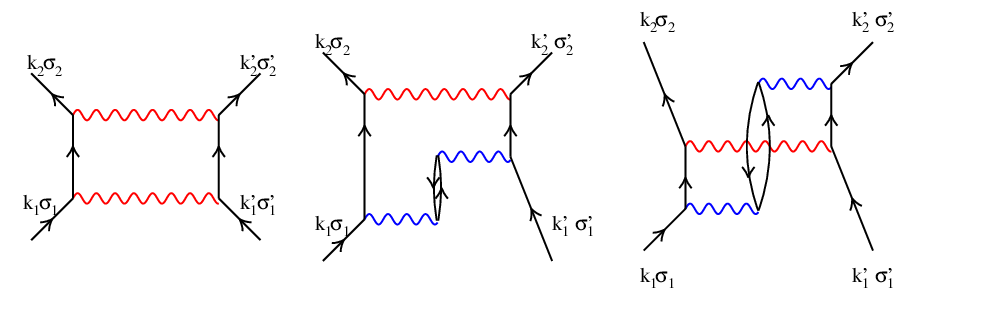}
    \caption{(color online) The figure shows the simplest second-order
      ladder diagrams including a ``twisted chain'' correction. The
      left diagram is the ordinary two-body ladder that is summed by
      the Bethe-Goldstone equation. The middle diagram is where one of the bare
      interactions is replaced by $\tilde w_I(q)$, and the right one is the simplest
      contribution to the totally irreducible interaction. The red
      wavy line represents the bare interaction and the blue wavy line
      the particle-hole interaction. The chains of two blue wavy lines
      may, of course, be supplemented by longer
      chains.}\label{fig:ladders}
\end{figure}

We have developed a systematic way to sum the beyond-parquet diagrams
in Ref. \onlinecite{v3twist}. The $G$ matrix of the Bethe-Goldstone
equation $\hat G(\rvec)$ is the sum of the ladder-diagram summation of
the bare interaction $\hat v(\rvec)$ {\rm plus\/} the induced
interaction $\hat w_I(\rvec)$. A similar $G$ matrix $\hat G_w(\rvec)$
can be defined for the local induced interaction $w_I(\rvec)$.  The
irreducible interaction is then obtained by solving the integral
equation
\begin{equation}
  \begin{split}
    &\hat V_{\rm I}(\qvec)=\\
    &-\frac{1}{2}\sum_{\alpha,\beta}
  \int \frac{d^3q'}{(2\pi)^3\rho}\left[\tilde G^{(\alpha)}(|\qvec-\qvec'|)
  -\tilde V_{\rm I}^{(\alpha)}(|\qvec-\qvec'|)\right]\times
  \\&\phantom{-\frac{1}{2}}\times\frac{\tilde G_w^{(\beta)}(q')}{2\tF(q')}
  \times\Tr_1
  \left[\hat O_\beta(a,1)\left[\hat O_\alpha(a,b),
        \hat O_\beta(1,b)\right]\right]\,,\label{eq:vi}
  \end{split}
\end{equation}
where $\tilde G(\qvec)$ is the Fourier transform of $G(\rvec)$.
The diagrams summed by this procedure are shown in Fig. \ref{fig:vn_twist}.

\begin{figure}
  \centering
    \includegraphics[width=0.95\columnwidth]{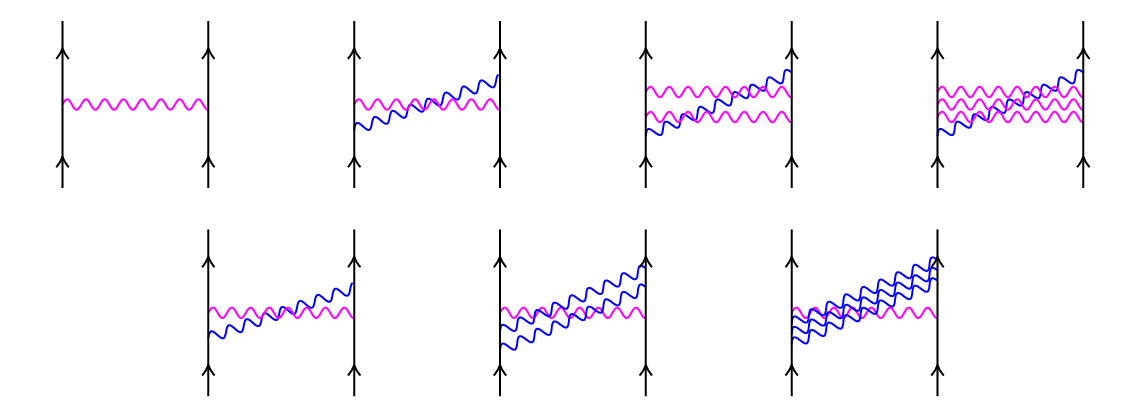}
    \caption{Examples of the diagrams summed by the integral equation
      \eqref{eq:vi}. The red wavy line represents a bare
      interaction $\hat v(r)$ and the blue line represents the sum
      $\hat v + \hat w_{\rm I}$.  The rungs can all be summed to the
      $G$-matrix.  }  \label{fig:vn_twist}
 \end{figure}

\subsection{Superfluid state with correlations}
\label{ssec:cbcs}

We rely in this section heavily on definitions and methods of
correlated basis functions (CBF) theory.  The basic idea of a
correlated BCS state is to use for the model state in
Eq. \eqref{eq:Jastrow} or \eqref{eq:f_prodwave} an uncorrelated BCS
state.
A {\em correlated\/} state is then constructed by applying a correlation
operator \eqref{eq:f_prodwave} to that state.
Since the superfluid state does not
have a fixed particle number, we must write the correlated state in
the form
\begin{equation}
\ket{\rm CBCS} =  \sum_{{\bf m},N} \ket {\Psi_{\bf m}^{(N)}}
\ovlp{{\bf m}^{(N)}}{\rm BCS}
\label{eq:CBCS}
\end{equation}
where the $\{\ket{{\bf m}^{(N)}}\}$ form a complete set of $N$-body Slater
determinants, and the $\ket {\Psi_{\bf m}^{(N)}}$ are correlated and normalized
$N$-body states forming a non-orthogonal basis of the Hilbert space.
\begin{equation}
  \ket {\Psi_{\bf m}^{(N)}} = \frac{F_N\ket{{\bf m}^{(N)}}}
       {\bra{{\bf m}^{(N)}}F_N^\dagger F_N \ket{{\bf m}^{(N)}}^{1/2}}\,.
\label{eq:IBCS}
\end{equation}

The approach to deal with triplet pairing that is closest to ours is
that of Hatzikonstantinou and Irvine \cite{IrvineBCS} who generalized
the work of Ref. \onlinecite{HNCBCS} to pairing in triplet states. The
uncorrelated wave function
\begin{equation}
  \ket{\rm{BCS}} = \prod_{\kvec\atop {k_x>0}}\left[
    u(\kvec) + \sum_{\sigma_1\sigma_2}v_{\sigma_1\sigma_2}(\kvec)\creat{\kvec,\sigma_1}\creat{-\kvec,\sigma_2}\right]\ket{}
  \label{eq:IrvineBCS}
  \end{equation}
is the superposition of particle pairs in a spin-triplet state
with opposite momenta. Above, $u(\kvec)$ is a normalization term.
    
In P-wave superconductivity \cite{PhysRev.131.1553}, the superfluid
system is frequently
generated by a generalized Bogoliubov
transformation
\begin{subequations}
  \label{eq:PBCS}
\begin{eqnarray}
  \ket{\rm BCS} &=& e^{\I S}\ket{},\\
  \I S &=& \frac{1}{2}\sum_{\kvec,\sigma_1,\sigma_2}\biggl[
      \theta(\kvec,\sigma_1,\sigma_2)
      \creat{\kvec,\sigma_1}\creat{-\kvec,\sigma_2} \nonumber\\ &&\qquad\quad
      - \theta^*(\kvec,\sigma_1,\sigma_2)\annil{-\kvec,\sigma_2}\annil{\kvec,\sigma_1}
      \biggr]\,.\end{eqnarray}
\end{subequations}
The above unitary transformation defines quasiparticle operators \cite{Tamagaki70}
\begin{equation}
  \balpha_{\kvec} = e^{\I S} {\bf a}_{\kvec} e^{-\I S} = {\bf U}(\kvec) {\bf a}_{\kvec}-{\bf V}(\kvec) {\bf a}^{\dagger}_{\kvec}
\end{equation}
where  ${\bf U}(\kvec)$ and ${\bf V}(\kvec)$ are 2$\times$2 matrices
and
\begin{equation}
  {\bf a}_{\kvec}
  = \begin{pmatrix}\annil{\kvec,\uparrow}\\ \annil{\kvec,\downarrow}\end{pmatrix}\,.\end{equation}
The superfluid ground state is then the state that is annihilated
by the quasiparticle destruction operators.  We hasten to note that
this state is not exactly the same as the state \eqref{eq:IBCS},
the states deviate by a 4-body term of the form
$\creat{\kvec\uparrow}\creat{-\kvec\uparrow}\creat{\kvec\downarrow}\creat{-\kvec\downarrow}$.
The relationship will be derived in Appendix \ref{app:unitary}.

The theory has been worked out in detail and applied to nuclear
problems by Tamagaki, Takatsuka, and collaborators
\cite{Tamagaki70,10.1143/PTP.48.1517}.

If the superfluid gap is small compared to the Fermi energy, it is
legitimate to simplify the problem by expanding
\begin{equation}
  \left\langle H'\right\rangle_c = \frac{\bra{\mathrm{CBCS}} \hat H'
    \ket{\mathrm{CBCS}}}
  {\bigl\langle\mathrm{CBCS}\big|\mathrm{CBCS}\bigr\rangle}\,,\qquad
  \hat H' \equiv \hat H - \mu\hat N\,
  \label{eq:EBCS}
\end{equation}
in the deviation of the superfluid values of $u(\kvec)$ and $v_{\sigma_1\sigma_2}(\kvec)$ from their normal states values $u^{(0)}(\kvec)=1-n(k)$,
 $v^{(0)}_{\sigma_1\sigma_2}(\kvec) = n(k)\delta_{\sigma_1\sigma_2}$.
Carrying out this expansion in the number of Cooper pairs, one arrives
\cite{HNCBCS,CBFPairing} at a gap equation of exactly the same form as
a mean-field approach, except that the pairing interaction is
expressed as a sum of FHNC or parquet diagrams.

For pairing in states other than $S$-waves, the gap function becomes
angle-dependent. We follow here the strategy of the ``angle-average''
approximation of the gap equation as formulated, for example, in Ref.
\onlinecite{Baldo3P23F2}. In general, the gap equation can couple
different angular momenta and becomes a matrix equation of the form
\begin{eqnarray}
  &&\Delta^{(\ell)}(k) =\label{eq:multigap}\\
  &&-\frac{1}{2}\sum_{\ell'}\int \frac{d^3k'}{(2\pi)^3}
  V_{\ell\,\ell'}(k,k')\frac{\Delta^{(\ell')}(k')}{
\sqrt{(\varepsilon_{k'}-\mu)^2 + D^2(k')}}\,.\nonumber
\end{eqnarray}
where $D^2(k) = \sum_\ell \left|\Delta^{(\ell)}(k)\right|^2$.
The pairing interaction has the form
\begin{eqnarray}
  V_{\ell,\ell'}(k,k') &=& {\cal W}_{\ell,\ell'}(k,k') \\
  &+&(|e_{\kvec}- \mu | + |e_{\kvec'}- \mu |)
{\cal N}_{\ell,\ell'}(k,k')
\label{eq:Pdef}\,,\nonumber\\
{\cal W}_{\ell,\ell'}(k,k') &=& \bra{k,\ell}
{\cal W}(1,2)\ket{k',\ell'}\,,\label{eq:Wnldef}\\
{\cal N}_{\ell,\ell'}(k,k')&=&
\bra{k,\ell}
{\cal N}(1,2)\ket{k',\ell'}\,.
\label{eq:Ndef}\end{eqnarray}
where we have suppressed spin degrees of freedom because we are always
working in either an $S=0$ or an $S=1$ state.  The effective
interaction ${\cal W}(1,2)$ and the correlation corrections ${\cal
  N}(1,2)$ are given by the compound-diagrammatic ingredients of the
FHNC-EL method for off-diagonal quantities \cite{CBF2}, or, in a
different language, by parquet-diagram summations.

\section{Results}
\label{sec:results}

Throughout our calculations, we have utilized a non-interacting
single-particle spectrum $\epsilon_k = \hbar^2 k^2/2m$. One can go
beyond such a simplifying approximation by either using the
single-particle spectrum predicted by correlated basis functions
theory \cite{HNCBCS} or improve upon that by including dynamic
effects.  These methods have been very successful for explaining the
physical mechanisms leading to the strong effective mass enhancement
in $^3$He \cite{he3mass} but applying them in the present case
appeared excessive.  The effective mass ratio $m^*/m$ lies between
1.05 and 0.95 \cite{ectpaper} and basically scales the gap, whereas,
as we will see, the effect we are reporting can change the gap by two
orders of magnitude.

As an initial exercise, and for completeness, we have solved the
$S$-wave and $P$-wave gap equations for the bare interactions.  We
used the Reid-68, the two operator versions of Reid-93, and the
Argonne $v_8$ potentials.  The results are discussed and compared with
earlier work in Appendix \ref{app:baregaps}.  To summarize the main
conclusions:
\begin{itemize} 
\item The results for the $^1$S$_0$ gap agree closely with each other,
  which gives confidence that the above interactions are well
  understood and leaves only many-body effects to influence the
  magnitude of the gap.  The temperature dependence is practically
  identical for the Reid-93 and Argonne $v_8$ potentials.
\item The results for the triplet pairing gap when many-body
  correlations are neglected are in full agreement with previous
  results \cite{KKC96,Baldo3P23F2,KhodelClark2001} {\em i.e.}, the
  diagonal $^3$P$_2$ and off-diagonal $^3$P$_2$-$^3$F$_2$ pairing
  channels prevail in neutron matter, and the tensor interaction is
  essential.  The temperature dependence is almost identical to that
  in the $^1$S$_0$ channel.
\item When the spin-orbit interaction is turned off, the $^3$P$_2$ and
  $^3$P$_2$-$^3$F$_2$ gaps disappear.  On the other hand, turning off
  the spin-orbit interaction leads to a significant $^3$P$_0$ gap for
  all interactions.
\end{itemize}
The question we address here is how many-body correlations modify the
pairing interactions and the above-mentioned results for the pairing
gaps.

\subsection{Effective interactions}
\label{ssec:Vph}

We have in sections \ref{sssec:chains}-\ref{sssec:twists}
formulated a version of the FHNC-EL or local parquet theory that
retains, in the summation of chain diagrams, only the simplest
exchange diagrams. This version has been dubbed FHNC-EL//0 or
parquet//0. More complicated exchange diagrams are also important and
routinely kept \cite{v3eos,v3twist}, we refer to our previous
publications \cite{v3eos,v3twist,v4} for details. 

Input to our calculations of pairing gaps are our pairing matrix
elements and the ``energy numerator corrections'' shown in
Eq. (\ref{eq:Pdef}).  The latter might seem unfamiliar; discussion of
the significance of this term is found in Refs.  \citenum{cbcs} and
\citenum{ectpaper}. Basically, the formulation
(\ref{eq:multigap})-(\ref{eq:Pdef}) amounts to a reformulation of the
gap equation in terms of the $T$-matrix as carried out, for example,
in Ref. \citenum{PethickSmith}. This reformulation is necessary to
guarantee that the gap equation \eqref{eq:multigap} has solutions in
the limit of a contact interaction.

The straightforward application of the parquet-diagram summations
would suggest taking, for the pairing interaction, the localized
effective interactions \eqref{eq:WindRPA} and \eqref{eq:VLSlocal}
which are made energy independent by the prescription
\eqref{eq:Wlocal}.  Since the gap is normally small compared to the
kinetic energy of a particle at the Fermi surface, it is more
appropriate to take these interactions at zero energy. We hasten to
mention that the difference of results obtained in this way is
negligible.

We focus in this work on P-wave pairing and assume an interaction
in the operator representation \eqref{eq:vop}. Examples of S-wave pairing can be seen in Appendix \ref{app:baregaps}.
We can restrict ourselves to the $T=1$ case, \ie the interaction
contains only the four odd-numbered operators in Eq. \eqref{eq:vop}.
For the pairing problem, we need a definite total spin S=0 or S=1.
The central components of the interaction are then
\begin{subequations}
\begin{eqnarray}
  v_{\rm c,0}(r) &=& v_c(r) - 3 v_\sigma(r)\,,\\
  v_{\rm c,1}(r) &=& v_c(r) +  v_\sigma(r)\,.
\end{eqnarray}
\end{subequations}
For $^3$P$_0$ pairing the interaction matrix \eqref{eq:Pdef} is
diagonal,
\begin{subequations}
  \label{eq:veffs}
\begin{eqnarray}
    &&V_{^3\mathrm{P}_0}(p_1,p_2) =\label{eq:V3P0}\\
    &&\int d^3r \left[v_{c,1}(r) - 4 v_{\rm S}(r)
    -2 v_{\rm LS}(r)\right]j_1(p_1 r) j_1(p_2 r)\,.\nonumber
  \end{eqnarray}
For $^3$P$_2$-$^3$F$_2$ pairing we obtain 
a $2\times 2$ matrix with
the elements
  \begin{eqnarray}
    &&V_{^3\mathrm{P}_2-^3\mathrm{P}_2}(p_1,p_2) =\label{eq:V3P2}\\
    &&\int d^3r \left[v_{c,1}(r) - \frac{2}{5} v_{\rm S}(r)
      + v_{\rm LS}(r)\right]j_1(p_1 r) j_1(p_2 r)\,,\nonumber\\
    &&V_{^3\mathrm{P}_2-^3\mathrm{F}_2}(p_1,p_2) =\label{eq;V3PF}\\
    &&\frac{6}{5}\sqrt{6}\int d^3r v_{\rm S}(r)
    j_1(p_1 r) j_3(p_2 r)\,,\nonumber\\
    &&V_{^3\mathrm{F}_2-^3\mathrm{F}_2}(p_1,p_2) =\label{eq:V3F2}\\
    &&\int d^3r \left[v_{c,1}(r) - \frac{8}{5} v_S(r)
      -4 v_{\rm LS}(r)\right]j_3(p_1 r) j_3(p_2 r)\,.\nonumber
  \end{eqnarray}
\end{subequations}
The effective interaction ${\cal W}(1,2)$ and the normalization
correction ${\cal N}(1,2)$ are originally represented in the same
operator basis as the bare interaction, their matrix elements entering
the gap equation are calculated according to Eqs.
\eqref{eq:V3P0}-\eqref{eq:V3F2}.

We show in Figs. \ref{fig:veffRe68} the effective interactions for the
$V_8$ version of the Reid 68 interaction in the $^3$P$_0$ and the
coupled $^3$P$_2$-$^3$F$_2$ coupled channels. Evidently there is not
much similarity between the bare interactions and the effective
interactions that contain medium-polarization, correlations, and
``twisted'' spin-exchange processes.

\begin{figure*}
  \centerline{\includegraphics[width=0.55\columnwidth,angle=-90]{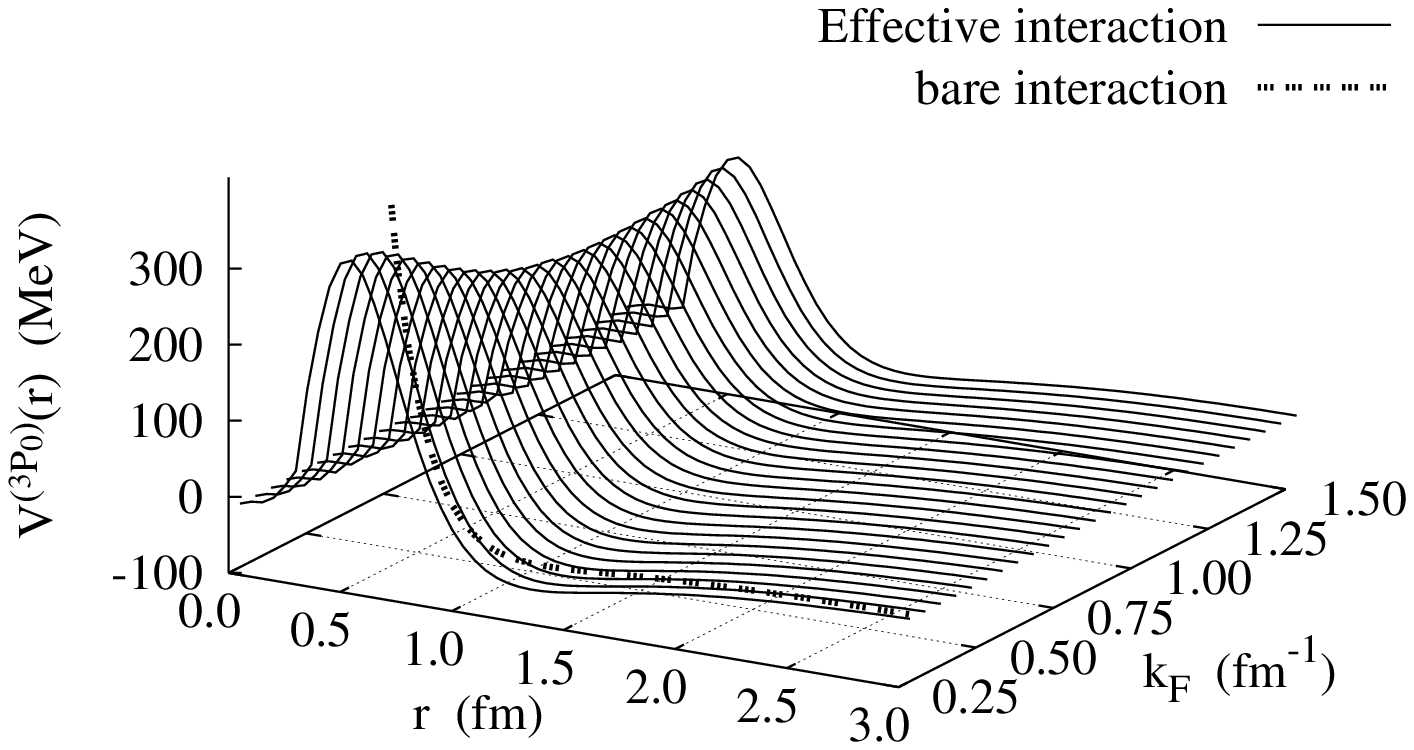}
    \includegraphics[width=0.55\columnwidth,angle=-90]{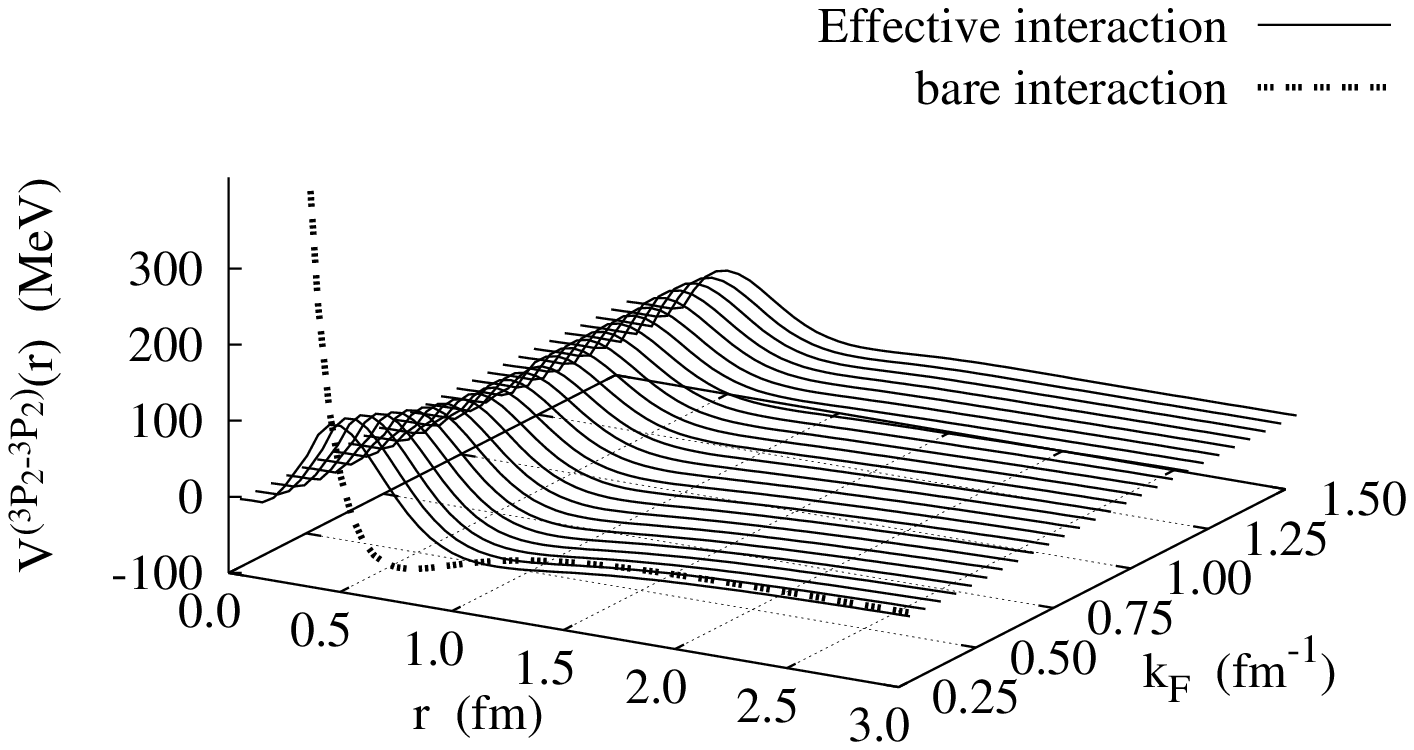}}
(a) \hspace{\columnwidth} (b)\\ 
  \centerline{\includegraphics[width=0.55\columnwidth,angle=-90]{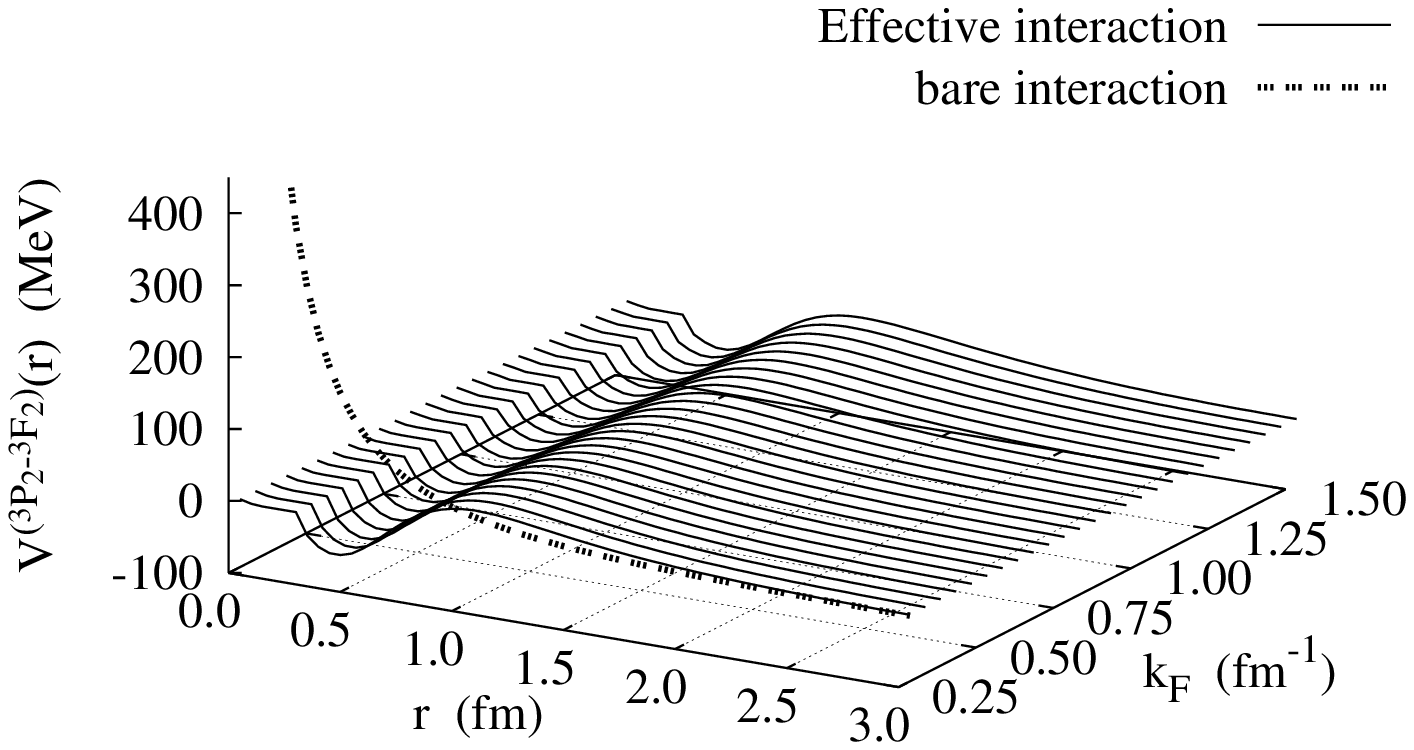}
  \includegraphics[width=0.55\columnwidth,angle=-90]{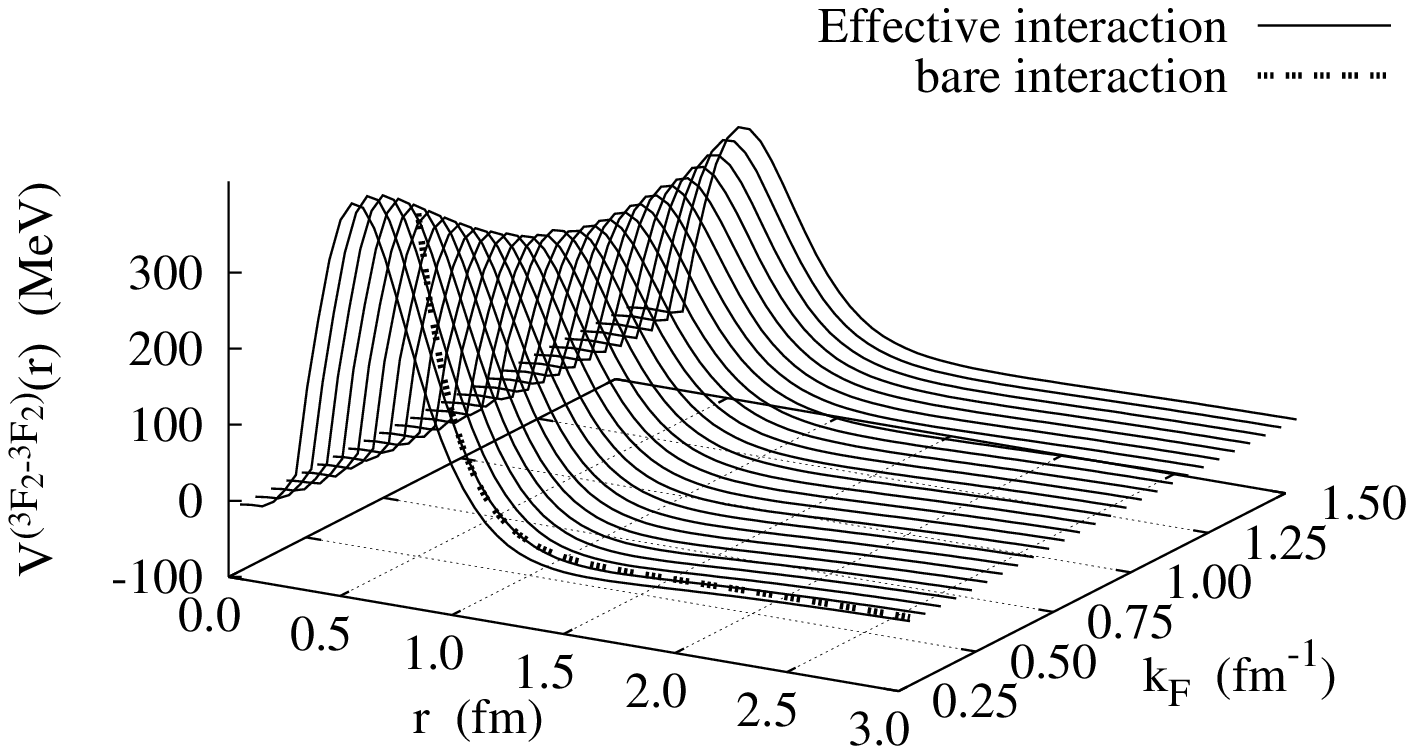}}
(c) \hspace{\columnwidth} (d) 
  \caption{The figures show, for the Reid $v_8$ interaction, the
    four effective interactions in (a) the $^3$P$_0$, the
    (b) the $^3$P$_2$-$^3$P$_2$,  (c) the $^3$P$_2$-$^3$F$_2$, and (d) the $^3$F$_2$-$^3$F$_2$ channels,
    for a sequence of Fermi wave number $\KF$ values (solid
    lines). Also shown are the bare interactions in the same channels
    (heavy dashed lines).
    \label{fig:veffRe68} }
\end{figure*}

We found in Ref. \onlinecite{v4} that the corrections to the effective
interactions $\widetilde W^{(\alpha)}(q;\omega)$ for $\alpha = 1, 3, 5$ due to the
spin-orbit potential are very small. The only term that is of first
order in the spin-orbit potential is $\tilde W^{\rm(7)}(q;\omega) =
\tilde W^{\LSsup}(q;\omega)$.  The most important result of that work
is, however, the screening of the short-ranged behavior of the
spin-orbit interaction by the correlations caused by the surrounding
particles, see Fig. \ref{fig:vLSRe68}.

\begin{figure}
  \includegraphics[width=0.65\columnwidth,angle=-90]{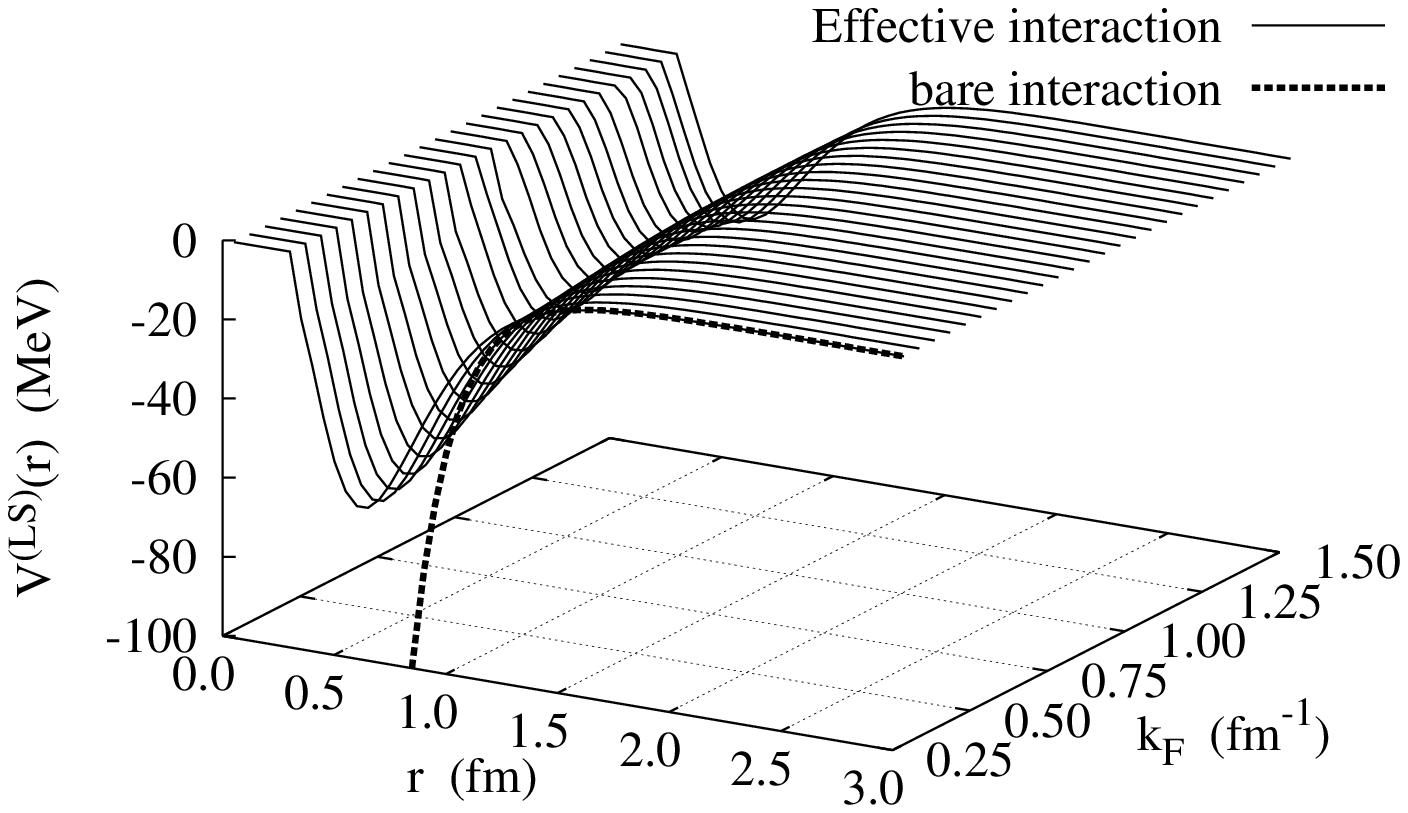}
  \caption{The figure shows, for the Reid $v_8$ interaction, the
    $W^{(\LSsup)}(r;\omega=0)$ for a sequence of Fermi wave number
    $\KF$. Also shown is the bare spin-orbit
    interaction (heavy dashed line).\label{fig:vLSRe68} }
\end{figure}

\begin{figure}
  \includegraphics[width=0.65\columnwidth,angle=-90]{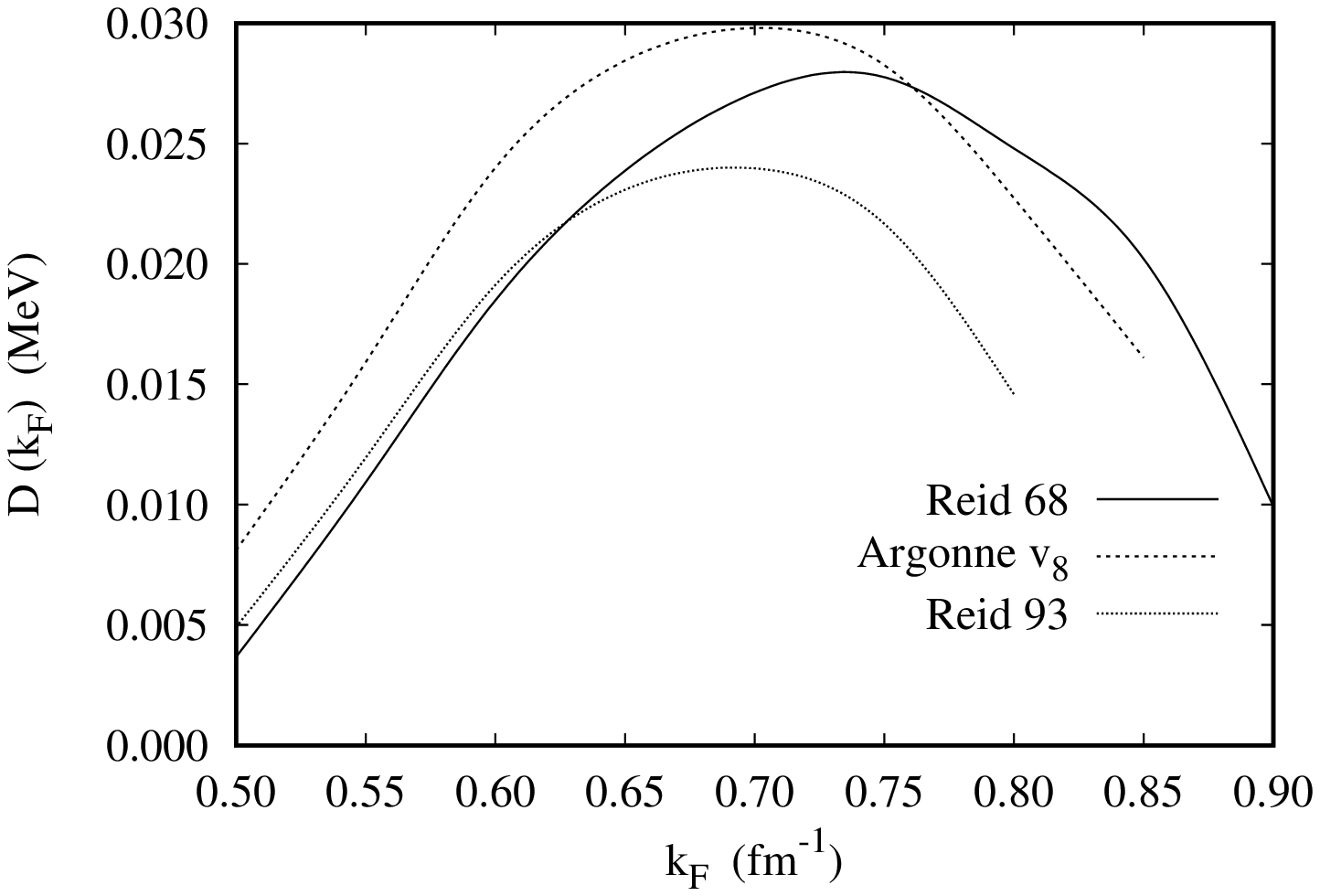}
\caption{The figure shows our results for the $^3$P$_2$-$^3$F$_2$  gap function
  $D(k)$ for the Reid 68, the Argonne $v_8$ and the Reid 93a interaction.
        \label{fig:3PFgaps}}
\end{figure}

\subsection{$^3$P$_2$-$^3$F$_2$ gaps}
\label{ssec:3P2gaps}

Fig. \ref{fig:3PFgaps} shows our final results for the superfluid
$^3$P$_2$-$^3$F$_2$ gap function $D(\KF)$. Since the gap is negligibly
small we did not dwell into technical details like the effect of
``twisted chain'' diagrams.


The most striking result is, of course, that, compared to the results
with the bare interactions (Fig.~\ref{fig:baregap3P2-3F2}), the
$^3$P$_2$-$^3$F$_2$ is suppressed by about two orders of
magnitude. The reason for this is readily found in the suppression of
the spin-orbit interaction through many-body correlations, as shown in
Fig.~\ref{fig:vLSRe68}.  The special role of the spin-orbit
interaction has already been pointed out in
Ref. \onlinecite{Gezerlis2014}: ``without an attractive spin-orbit
interaction, neutrons would form a $^3$P$_0$ superfluid, in which the
spin and orbital angular momenta are anti-aligned, rather than the
$^3$P$_2$ state, in which they are aligned.''

\subsection{$^3$P$_0$ gaps}
\label{ssec:3P0gaps}

Fig. \ref{fig:3P0gaps} shows our results for the superfluid $^3$P$_0$
gap, omitting and including ``beyond-parquet'' corrections.  We see a
similar effect as observed in our work on S-wave pairing but with the
opposite sign: Adding the ``twisted chain'' diagrams increases the
P-wave gap by about 30 percent. This is plausible considering our
discussion around Fig. \ref{fig:ladders}. Adding ``non-parquet''
diagrams in the spin-singlet channel mixes the repulsive P-wave
interaction into the S-wave pairing interaction. On the other hand,
adding ``non-parquet'' diagrams in the spin-triplet channel mixes the
attractive spin-singlet interaction to the spin-triplet pairing
potential. We hasten to point out the effect is, however, 
rather delicate since the suppression of the spin-orbit
potential depends sensitively on the balance between the central force
and the spin-orbit force in the spin-triplet channels.
It is less than the uncertainty introduced by using different microscopic
potentials.

\begin{figure*}
  \includegraphics[width=0.65\columnwidth,angle=-90]%
      {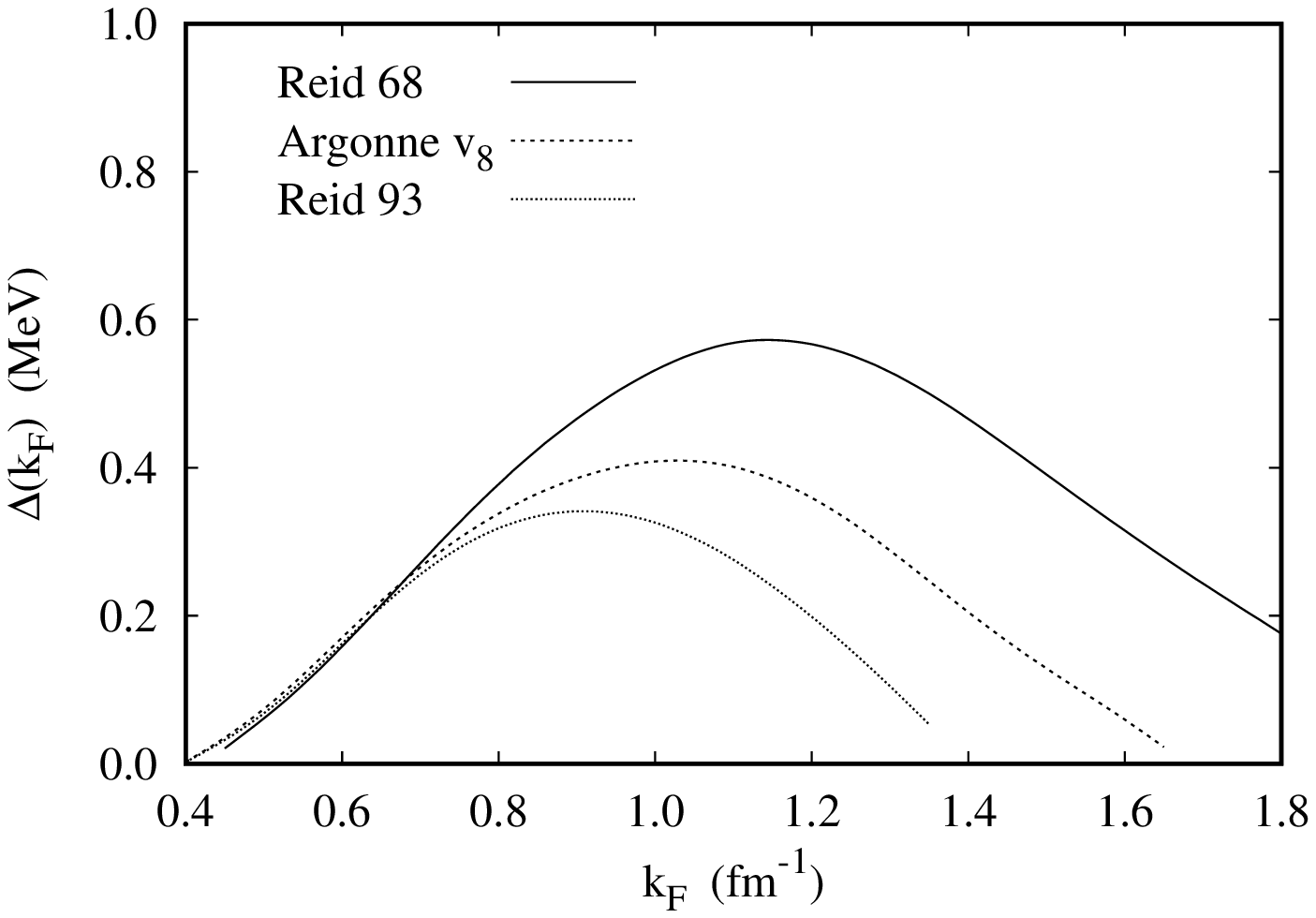}
  \includegraphics[width=0.65\columnwidth,angle=-90]%
      {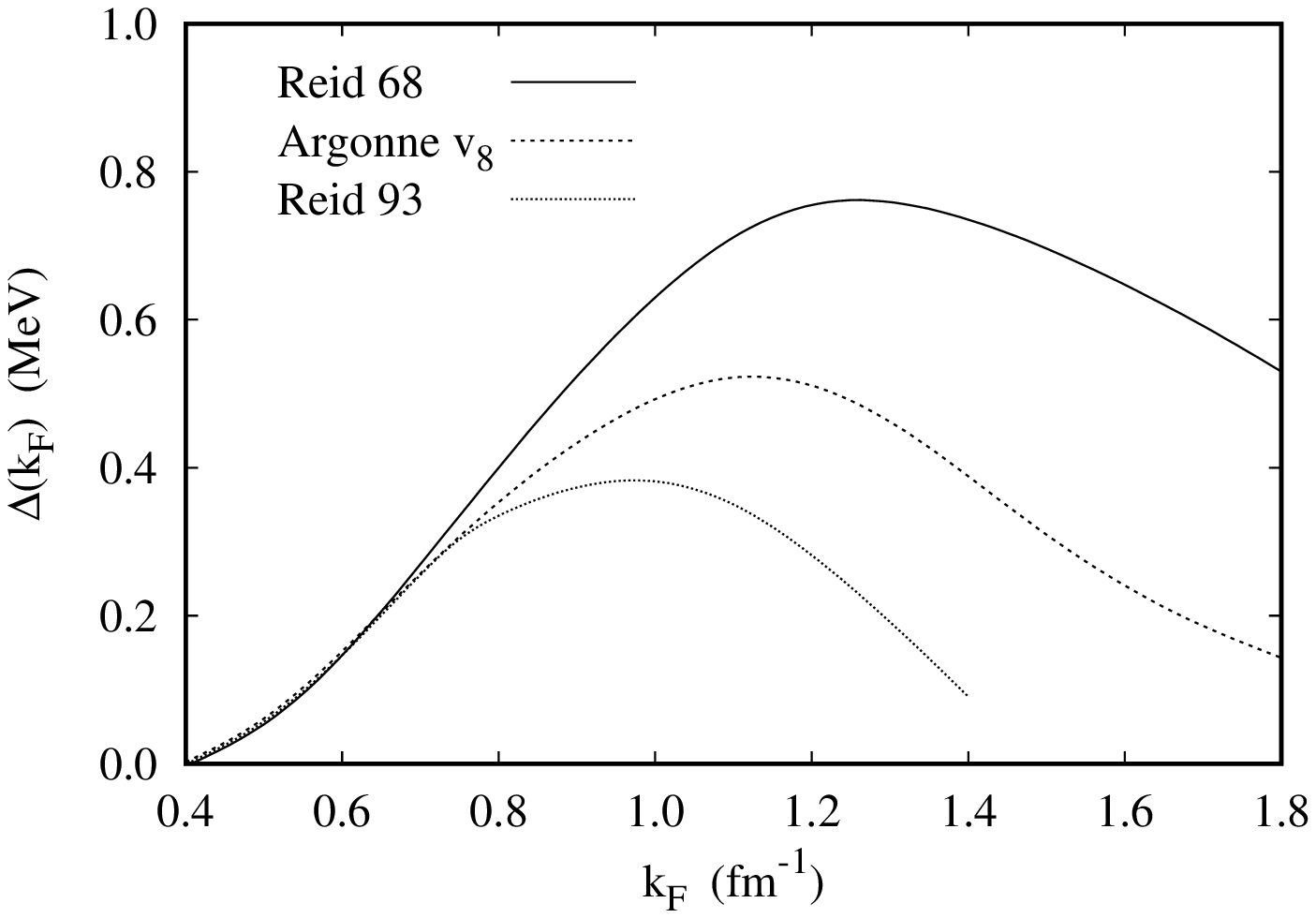}\\
(a) \hspace{0.8\columnwidth} (b)
      \caption{For the three potentials
        considered here, the $^3$P$_0$ gap is shown when (a) the ``beyond parquet''
        diagrams are omitted and (b) when they are included.
        \label{fig:3P0gaps}}
\end{figure*}
  
\begin{figure}
  \centerline{\includegraphics[width=0.7\columnwidth,angle=270]%
    {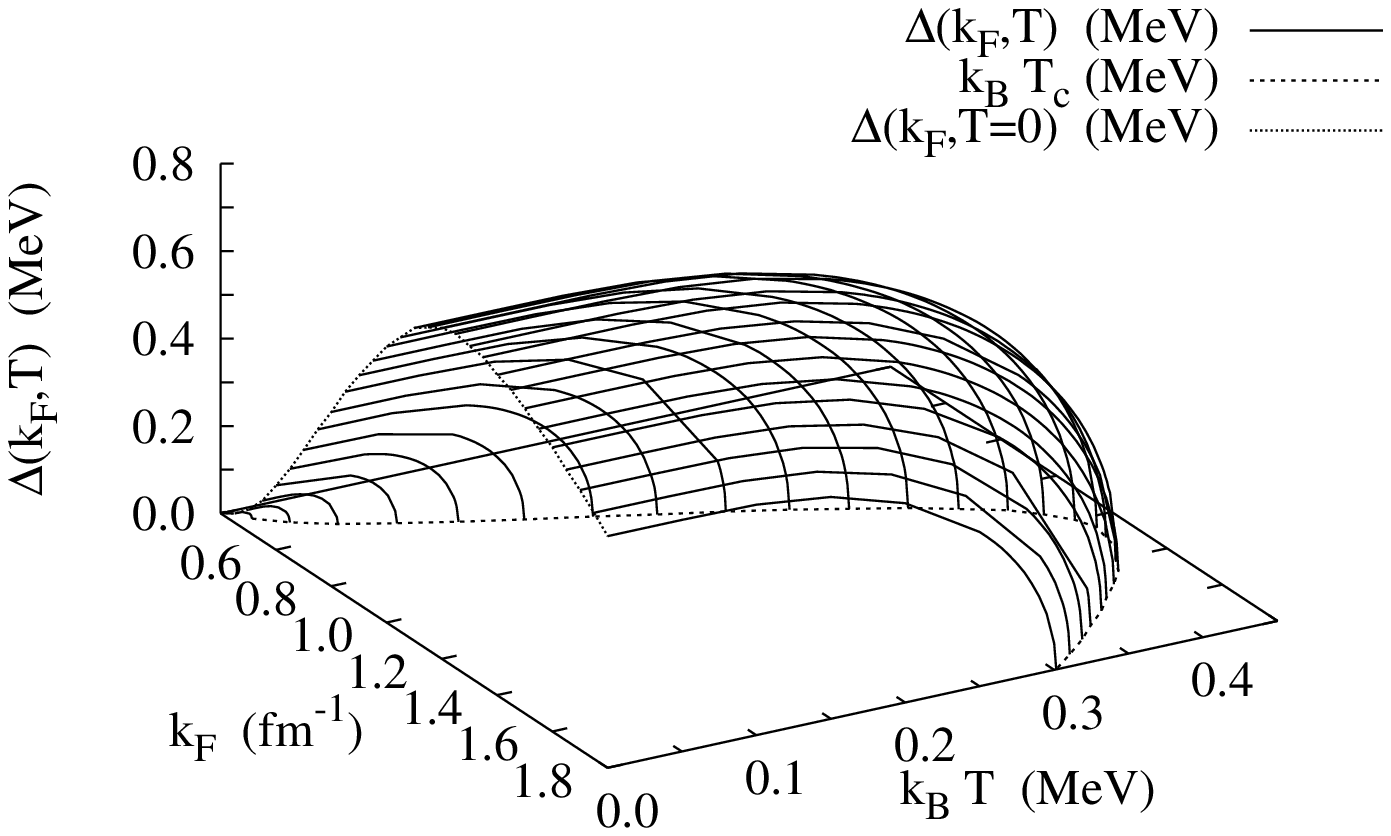}}
\caption{The figure shows, for the Reid 68 interaction, the
  temperature dependence of the $^3$P$_0$ gap. Also shown are the
  critical temperature $T_c$ as a function of density (dashed line in
  the $(\KF,\KB T)$ plane) and the zero temperature solution (dotted
    line in the $(\KF,\Delta(\KF))$ plane).  \label{fig:re68gapL93P0}}
\end{figure}

The fact that the ``correlated wave function'' {\em ansatz\/} \eqref{eq:CBCS}
leads to a description of a superfluid system that can be understood
as a weakly interacting theory with effective interactions permits us to
generalize the description to finite temperatures. The gap equation then
takes the form \cite{Schrieffer1999}
\begin{eqnarray}
  \Delta^{(\ell)}(k) &=&
  -\frac{1}{2}\sum_{\ell'}\int \frac{d^3k'}{(2\pi)^3}
  V_{\ell\,\ell'}(k,k')
   \nonumber \\  
 & &  \frac{\tanh\left(\frac{1}{2}\beta E(k')\right)}{E(k')}
  \Delta^{(\ell')}(k') \, , 
  \label{eq:gapT}
\end{eqnarray}
where $\beta = 1/\KB T$ and $E(k) = \sqrt{(\epsilon_{k}-\mu)^2 + D^2(k)}$.
  Fig. \ref{fig:re68gapL93P0} shows the temperature dependence of
  the $^3$P$_0$ gap for the Reid 68 interaction. In general, we find
  the relationship $\Delta(\KF) \approx 1.8\KB T_c$ in very good agreement
  with the mean-field estimate  $\Delta(\KF) \approx 1.76\KB T_c$
  \cite{FetterWalecka},
  although the relationship
  \begin{equation}
    \Delta(\KF,T) \sim \left(1-\frac{T}{T_c}\right)^{\frac{1}{2}}\,.
    \label{eq:DofT}
  \end{equation}
  is satisfied only close to the critical temperature $T_c$.

\section{Summary and outlook}
\label{sec:summary}

Using several $v_8$ models of the
nucleon-nucleon interaction, 
we have calculated the effective pairing interaction in
neutron matter by summing the parquet-diagrams and important totally
irreducible corrections. In doing so, we have carried out the most
comprehensive diagrammatic evaluation of the pairing interactions,
including, among others, medium polarization effects and spin-flip
processes that have, so far, been either ignored or treated at a
simplistic phenomenological level.

One can certainly do better. In particular, at high densities, more
complicated exchange contributions can be relevant. These are
routinely included in systems with simple interactions \cite{polish},
but the summation techniques have not yet been developed for
state-dependent potentials. Therefore, we have decided not to go
beyond $\KF = 1.8\,\mathrm{fm}^{-1}$.

The most striking result is that many-body correlations have the
effect of almost completely suppressing the $^3$P$_2$ and
$^3$P$_2$-$^3$F$_2$ gap. This result was, to some extent, anticipated
from our work on the spin-orbit interaction \cite{v4}. There, we
found that the spin-orbit interaction is strongly suppressed by
many-body correlations. It is also consistent with the fact that
neither of the {\rm bare\/} interactions employed here shows
$^3$P$_2$-$^3$F$_2$ or $^3$P$_2$ pairing when the spin-orbit
interactions is omitted.  Along with the suppression of
$^3$P$_2$-$^3$F$_2$ pairing we found a significantly enhanced
pairing in $^3$P$_0$ states.

There are a number of obvious ways of extending our calculations.  One
of them is the inclusion of three-nucleon forces.  Generally, the
combined effect of 3N forces and in-medium mass normalization can vary
strongly.  A treatment of the three-nucleon interaction adjusted for
applications in Brueckner-Hartree-Fock theory leads to a moderate
increase near the peak \cite{PhysRevC.69.018801}.  On the other hand,
the depletion of the Fermi surface can drastically reverse the
effect~\cite{DLZ2013}.  In Ref.~\onlinecite{JLTP189_361}, it was found that
a small fraction of the phenomenological repulsion of the original
Urbana potential suffices to eliminate the gap. 
In the case of $\chi$EFT potentials, the contribution of the three-nucleon force at N$^2$LO and N$^3$LO  is somewhat attractive, 
leading to an enhancement of the gap which is strongly regulator-dependent \cite{JLTP189_234,Drischler2017}, 
especially beyond Fermi momentum $1.5~$fm$^{-1}$.  
Note that such momentum scales are close to the breakdown scale of $\chi$EFT.   

Another potential extension of our work is to use a superfluid
Lindhard function \cite{PhysRevB.61.9095} for the calculation of the
induced interaction. The effect can be quite large
\cite{fullbcs,v3bcs} in low-density neutron matter where the gap can
be as large as half of the Fermi energy. We have not included these
corrections here because the $^3$P$_0$ gap is much smaller than the
$^1$S$_0$ gap and has its maximum at higher densities. In view of the
approximations implicit to our parquet//1 calculation, we did not
consider the effort justified.

\appendix
\begin{widetext}
\section{Unitary Transformation}
\label{app:unitary}

The Bogoliubov unitary transformation approach introduces a unitary
operator \cite{BeliaevLesHouches,Tamagaki70}

\begin{equation}
  \I S = \frac{1}{2}\sum_{\kvec,\sigma_1,\sigma_2}
  \left[\theta_{\sigma_1\sigma_2}(\kvec)\creat{\kvec\sigma_1}
    \creat{-\kvec\sigma_2}-\theta^*_{\sigma_1\sigma_2}(\kvec)
    \annil{-\kvec\sigma_2}\annil{\kvec\sigma_1}\right]\nonumber\\
  \equiv\frac{1}{2}\sum_\kvec \I s(\kvec)
  \label{eq:UnitaryBCS}\,,\end{equation}
and, through that, a set of quasiparticle operators
\begin{equation}
  \balpha_{\kvec} = e^{\I S}{\bf a}_{\kvec} e^{-\I S}
    = {\bf U}(\kvec){\bf a}_{\kvec} - {\bf V}(\kvec){\bf a}^\dagger_{-\kvec}
\end{equation}
where
\begin{equation}
  {\bf a}_{\kvec} = \begin{pmatrix}\creat{\kvec\uparrow}\\
      \creat{\kvec\downarrow}\\\end{pmatrix}\label{eq:QPdef}
\end{equation}
and ${\bf U}(\kvec)$ and ${\bf V}(\kvec)$ are $2\times 2$ matrices.

In the spin-triplet case, $\theta_{\sigma_1\sigma_2}(\kvec)$ is
symmetric, and we have
$\theta_{\sigma_1\sigma_2}(\kvec)=-\theta_{\sigma_1\sigma_2}(-\kvec)$
as well as \cite{10.1143/PTP.48.1517}
$\theta^*_{-\sigma_1-\sigma_2}(\kvec) = (-1)^{\sigma_1+\sigma_2}
\theta_{\sigma_1\sigma_2}(\kvec)$, and we can write
\begin{equation}
  \I S= \sum_{\kvec \atop {k_z>0}}\I s(\kvec)\,.
  \end{equation}
Then all $s(\kvec)$ with different $\kvec$ in the above sum commute.
The matrices ${\bf U}(\kvec)$ and ${\bf V}(\kvec)$
have the form \cite{10.1143/PTP.48.1517}
\begin{equation}
  {\bf U}(\kvec) = \begin{pmatrix}
    u(\kvec) & 0 \\ 0 & u(\kvec)\\
    \end{pmatrix}\quad{\bf V}(\kvec) = \begin{pmatrix}
    v(\kvec) & 0 \\ 0 & v(\kvec)\\
  \end{pmatrix}{\bm \theta}(\kvec)
  \end{equation}
  where $u(\kvec) = \cos\theta_D(\kvec)$,  $v(\kvec) = \sin\theta_D(\kvec)/\theta_D(\kvec)$ and $\theta_D(\kvec) = \sqrt{\left|\theta_{\uparrow\uparrow}(\kvec)
    \right|^2 + \left|\theta_{\uparrow\downarrow}(\kvec)
    \right|^2}$.
  Conventionally, the Hamiltonian is rewritten in terms of the
    quasiparticle operators, the amplitudes $u(\kvec)$ and $\theta_{\sigma_1\sigma_2}(\kvec)$ are determined by the condition that the off-diagonal
    part of the Hamiltonian in terms of the quasiparticle operators
    vanishes.
    For the execution of the variational/parquet theory outlined in section
    \ref{ssec:cbcs} we need, however, a closed-form expression for
    the uncorrelated state $\ket{\mathrm{BCS}}$. One definition
    is to demand that this state is destroyed by the quasiparticle
    annihilation operators,
    \begin{equation}
      \balpha_{\kvec}\ket{\mathrm{BCS}} = 0\,.
    \end{equation}
It is immediately seen that such a condition can not be satisfied
by the form \eqref{eq:IBCS} and the quasiparticle operators
    \eqref{eq:QPdef}. Rather, the form of the wave function is
    \begin{eqnarray}
      \ket{\mathrm{BCS}} &=& \prod_{\kvec\atop{k_z>0}} F(\kvec)\ket{}\\
      F(\kvec) &=& A(\kvec) + \sum_{\sigma_1\sigma_2}B_{\sigma_1\sigma_2}(\kvec)\creat{\kvec\sigma_1}
    \creat{-\kvec\sigma_2}
        + C(\kvec)\creat{\kvec\uparrow}\creat{-\kvec\uparrow}\creat{\kvec\downarrow}\creat{-\kvec\downarrow}\end{eqnarray}
where the coefficients $A(\kvec)$, $B_{\sigma_1\sigma_2}(\kvec)$ and
$C(\kvec)$ are determined by the condition
    \begin{equation}
    \sum_{\sigma'}\Bigl[U_{\sigma,\sigma'}(\kvec)\annil{\kvec,\sigma'}
      - V_{\sigma,\sigma'}(\kvec)\creat{-\kvec,\sigma'}\Bigr]F(\kvec)\Ket{}
  \end{equation}
    and normalization.
The first condition is that there are no terms proportional to $\creat{-\kvec,\sigma}$. This leads to
    \begin{equation}
      u(\kvec)B_{\sigma\sigma'}(\kvec) = A(\kvec)V_{\sigma\sigma'}(\kvec)\,.\label{eq:first}
    \end{equation}
 The  second condition is that the coefficients of three creation operators
 should also vanish.
  \begin{eqnarray}
    &&-C(\kvec)\left[U_{\sigma\uparrow}\creat{\kvec,\downarrow}-U_{\sigma\downarrow}
      \creat{\kvec,\uparrow}\right]
    -\sum_{\sigma_1}\left[V_{\sigma\uparrow}(\kvec)B_{\sigma_1\downarrow}(\kvec)
      - V_{\sigma\downarrow}(\kvec)B_{\sigma_1\uparrow}(\kvec)\right]\creat{\kvec,\sigma_1}\nonumber\\
    &=&-\left[u(\kvec)C(\kvec)-
     \left[V_{\uparrow\uparrow}(\kvec)B_{\downarrow\downarrow}(\kvec)-
       V_{\uparrow\downarrow}(\kvec)B_{\downarrow\uparrow}(\kvec)\right]\right]
\left[\delta_{\sigma\uparrow}\creat{\kvec,\downarrow}-\delta_{\sigma\downarrow}
  \creat{\kvec,\uparrow}\right]\,.
    \end{eqnarray}
  Together with the normalization condition $\bra{}F^\dagger(\kvec)F(\kvec)\ket{}=1$
  this gives the result
    \begin{equation}
      F(\kvec) = u^2(\kvec) + u(\kvec)\sum_{\sigma_1\sigma_2}V_{\sigma_1\sigma_2}(\kvec)\creat{\kvec\sigma_1}
    \creat{-\kvec\sigma_2}
    + (u^2(\kvec)-1)\creat{\kvec\uparrow}\creat{-\kvec\uparrow}\creat{\kvec\downarrow}\creat{-\kvec\downarrow}\,.\end{equation}
With that, the wave function is
\begin{eqnarray}
\Ket{\rm{BCS}} &=& \prod_{\kvec\atop k_z>0}
\biggl[u^2(\kvec) + u(\kvec)\sum_{\sigma_1,\sigma_2}
 V_{\sigma_1,\sigma_2}(\kvec)
 \creat{\kvec,\sigma_1}\creat{-\kvec,\sigma_2}\nonumber\\
 &+&
 (u^2(\kvec)-1)\creat{\kvec\uparrow}\creat{-\kvec\uparrow}\creat{\kvec\downarrow}\creat{-\kvec\downarrow}\biggr]\ket{}\,.
\end{eqnarray}
Thus, the ``unitary transformation'' leads to a BCS state that also contains
four creation operators. Such a term is, of course, irrelevant if the
Hamiltonian just contains one- and two-body operators. It might lead to interesting effects once one goes beyond mean-field approximations.

\section{Results with bare interactions.}
\label{app:baregaps}

To examine the dependence of the results for the superfluid gap on the
potential model and also to compare with earlier calculations
\cite{KKC96,Baldo3P23F2,KhodelClark2001}, we have calculated
$\Delta(\KF)$ for the four potentials studied here. An aspect of
concern is that most accurate nuclear interactions
\cite{Reid68,Reid93} are given in a partial wave basis, and an
operator form (\ref{eq:vop}, \ref{eq:operator_v8}) is an approximation.
We have described above how the operator form of the Reid-93
interaction was obtained.

Fig. \ref{fig:baregaps1S0} shows a comparison of the $^1$S$_0$ gap
obtained for the above three interaction models. The close agreement
between these results gives confidence that these interactions are
well understood, leaving only many-body effects to influence the
magnitude of the gap.

Extending the calculation to finite temperature
\cite{MorseBohm1957,FetterWalecka,Schrieffer1999} also gives an estimate
of the critical
temperature $T_c$. Close to the transition temperature, one expects
a behavior of the form \eqref{eq:DofT}.

We have calculated the temperature dependence of the $^1$S$_0$ gap,
see Fig. \ref{fig:re68gap1S0T}; the critical temperature was obtained
by extrapolating the temperature dependence $\Delta(\KF,T)$ to
$\Delta(\KF,T_c)=0$ using the estimate \eqref{eq:DofT}. Our results
are shown in Fig. \ref{fig:re68gap1S0T}.  We found, to a very good
approximation that $\Delta(\KF) \approx 1.8\KB T_c$ throughout the
whole density regime. This is in excellent agreement with the weak
coupling approximation Eq. (51.44) of Ref. \onlinecite{FetterWalecka}.
Results for the Argonne and the Reid 93 interaction are practically
identical.

The situation is more complex for P-wave pairing. The pioneering
work of Tamagaki \etal \citep{Tamagaki70,10.1143/PTP.48.1517} showed
that, when many-body effects are neglected, $^3$P$_2$ and
$^3$P$_2$-$^3$F$_2$ prevail in neutron matter. The results that we
have obtained for the interactions considered here fully support this
view.  Moreover, our results are rather similar for all four cases, see
Fig, \ref{fig:baregap3P2-3F2}. We also found that, in the
$^3$P$_2$-$^3$F$_2$ coupled channel pairing, the interactions in the
$^3$P$_2$-$^3$P$_2$ diagonal and the $^3$P$_2$-$^3$F$_2$ off-diagonal
(tensor) channel are most important whereas the diagonal interaction
$^3$F$_2$-$^3$F$_2$ interaction plays only a minor role. On the other hand,
the off-diagonal tensor force is essential; restricting the
calculation to the $^3$P$_2$ channel reduces the gap by an order
of magnitude, see Fig. \ref{fig:baregap3P2}.

The temperature dependence of the $^3$P$_2$-$^3$F$_2$ gap is almost
identical to that of the $^1$S$_0$ gap, see
Fig. \ref{fig:re68gap3P2-3F2T}. In particular, we found again that the
relationship $\Delta(\KF) \approx 1.8\KB T_c$ with very good accuracy.

The calculations discussed so far did not necessarily rely on an
operator structure of the form (\ref{eq:vop},\ref{eq:operator_v8}).
To assess the importance of the spin-orbit interaction we have
repeated our calculations with the spin-orbit interaction turned off.
As predicted \cite{Gezerlis2014}, pairing in $^3$P$_2$-$^3$F$_2$ and
$^3$P$_2$ states disappeared apart from a tiny effect of $10^{-2}$MeV
for $^3$P$_2$ pairing with the Reid 68 interaction.

On the other hand, turning off the spin-orbit interaction leads to a
significant $^3$P$_0$ gap for all interactions considered here, see
fig. \ref{fig:baregaps3P0_noLS}. Unlike all other cases, we observe a
rather significant dependence of the value of $\Delta(\KF)$ on the
interaction. This simply reflects the ambiguity arises from the very definition of
an operator structure of the interaction.
In particular, the predictions of the two operator representations
\eqref{eq:R93a} and \eqref{eq:R93b} of the Reid 93 interaction differ
by more than a factor of 2.

\section{Gap equation solver}
\label{app:gapsolver}

We describe here the version of our gap-equation solver \cite{EKDiplom}
for finite temperatures.
For brevity and in view of what follows, let
\begin{equation}
E_n(k,\xi) \equiv \sqrt{(\varepsilon_{k}-\mu)^2 + \xi^2 D_n^2(k)}\,.
\end{equation}
The equations \eqref{eq:gapT} and \eqref{eq:multigap} are highly non-linear and a simple
iteration procedure of the kind
\begin{equation}
  \Delta^{(\ell)}_{n+1}(k) =
  -\frac{1}{2}\sum_{\ell'}\int \frac{d^3k'}{(2\pi)^3}
  V_{\ell\,\ell'}(k,k')
  \frac{\tanh\left(\frac{1}{2}\beta E_n(k',1)\right)}{E_n(k',1)}
  \Delta_n^{(\ell')}(k')
\label{eq:gapiter}
\end{equation}
normally does not converge. The reason for this is as follows: The
function
\[\frac{\tanh\left(\frac{1}{2}\beta E_n(k',1)\right)}
       { E_n(k',1)}\]
is everywhere apart
  from a small area around $\KF$ dominated by the kinetic energy
  $|\varepsilon_{k'}-\mu|$.  We can therefore write (applying the mean
  value theorem)
\begin{equation}
  \Delta^{(\ell)}_{n+1}(k) =
  -\frac{1}{2}\sum_{\ell'}\int \frac{d^3k'}{(2\pi)^3}
  V_{\ell\,\ell'}(k,k')
  \frac{\tanh\left(\frac{1}{2}\beta\sqrt{(\varepsilon_{k'}-\mu)^2 + \bar D^2(\KF)}\right)}{
\sqrt{(\varepsilon_{k'}-\mu)^2 + \bar D^2(\KF)}}\Delta^{(\ell')}_{n}(k')
\end{equation}
where $\bar D(\KF)$ is an average value of the gap function
$D(k)$ in the vicinity of $\KF$.  This procedure converges towards the
eigenvector corresponding to the {\em largest eigenvalue in absolute value
  \/} of the matrix
\[-\frac{1}{2}V_{\ell\,\ell'}(k,k')
\frac{\tanh\left(\frac{1}{2}\beta\sqrt{(\varepsilon_{k'}-\mu)^2 + \bar D^2(\KF)}\right)}{
  \sqrt{(\varepsilon_{k'}-\mu)^2 + \bar D^2(\KF)}}\,.\]

However, if the interaction has a strong repulsive core, the above
operator has very large {\em negative\/} eigenvalues whereas we {\em want\/}
the largest positive eigenvalue.

A very rapidly converging algorithm \cite{EKDiplom} is as follows:

Consider the generalized eigenvalue problem
\begin{equation}
  \lambda(\xi)\delta^{(\ell)}_{n+1}(k,\xi) = -\frac{1}{2}\sum_{\ell'}
  \int \frac{d^3k'}{(2\pi)^3}
  V_{\ell\,\ell'}(k,k')
  \frac{\tanh\left(\frac{1}{2}\beta E_n(k',\xi)\right)}
       {E_n(k',\xi)}\delta^{(\ell')}_{n+1}(k',\xi)
\label{eq:eigengap}
\end{equation}
where $\xi$ is a scaling parameter.  The eigenvectors are normalized as
\begin{equation}
  \sum_{\ell}\int \frac{d^3k}{(2\pi)^3}
  \displaystyle\frac{\frac{1}{2}\tanh\left(\frac{1}{2}\beta E_n(k,\xi)\right)}
{E_n(k,\xi) }\left|\delta^{(\ell)}_{n+1}(k)\right|^2= 1\,.\label{eq:norm}
\end{equation}

The algorithm is
\begin{enumerate}
\item{} Start with a reasonable estimate for $D_0(k)$. From the above
  analysis we can expect that a constant is a good approximation.
\item{} Solve the above eigenvalue problem as a function of the scaling
parameter $\xi$, find the value $\xi_0$ for which an eigenvalue
of the equation is $\lambda(\xi_0)=1$. The derivative is
\begin{eqnarray}
\frac{d\ln \lambda(\xi)}{d\xi}
&=&\sum_{\ell}\int \frac{d^3k}{(2\pi)^3}
\left|\delta^{(\ell)}_{n+1}(k)\right|^2
\frac{d}{d\xi}\frac{\tanh\left(\frac{1}{2}\beta E_n(k,\xi)\right)}
     {E_n(k,\xi) }\nonumber\\
     &=&-\xi \sum_{\ell}\int \frac{d^3k}{(2\pi)^3}
     \left|\frac{D_n(k)\delta^{(\ell)}_{n+1}(k)}{E_n(k,\xi)}\right|^2
     \frac{\tanh(\frac{1}{2}\beta E_n(k,\xi))}
     {E_n(k,\xi)}
     \left[1-\frac{\beta E_n(k,\xi)}{\sinh(\beta E_n(k,\xi))}\right]<0
\label{eq:ddelta}
\end{eqnarray}
for $\lambda > 0$. This feature is useful to find the value $\xi_0$ by
a Newton procedure.

\item{} Scale the corresponding eigenfunction
  $\delta^{(\ell)}_{n+1}(k,\xi_0)$ such that
  \[D_{n+1}(k) = \delta_{n+1}(k,\xi_0)
  \frac{\xi_0 D_n(\KF)}{\delta_{n+1}(\KF,\xi_0)}\,.\]
\item{} Go to step (2) and repeat until convergence which is reached
  for $\xi_0 = 1$.
\end{enumerate}
It was already observed in Ref. \onlinecite{EKDiplom} and confirmed in
Ref.  \onlinecite{Baldo3P23F2} that the first iteration often leads to
a solution of the gap equation with a percent accuracy. We have
confirmed this observation here. A stand-alone code together with a
brief description on how it works is provided as supplemental material
\cite{v43p2suppl}.
\end{widetext}

\mbox{~}
\newpage 
\clearpage 

\begin{figure}
  \centerline{\includegraphics[width=0.6\columnwidth,angle=270]{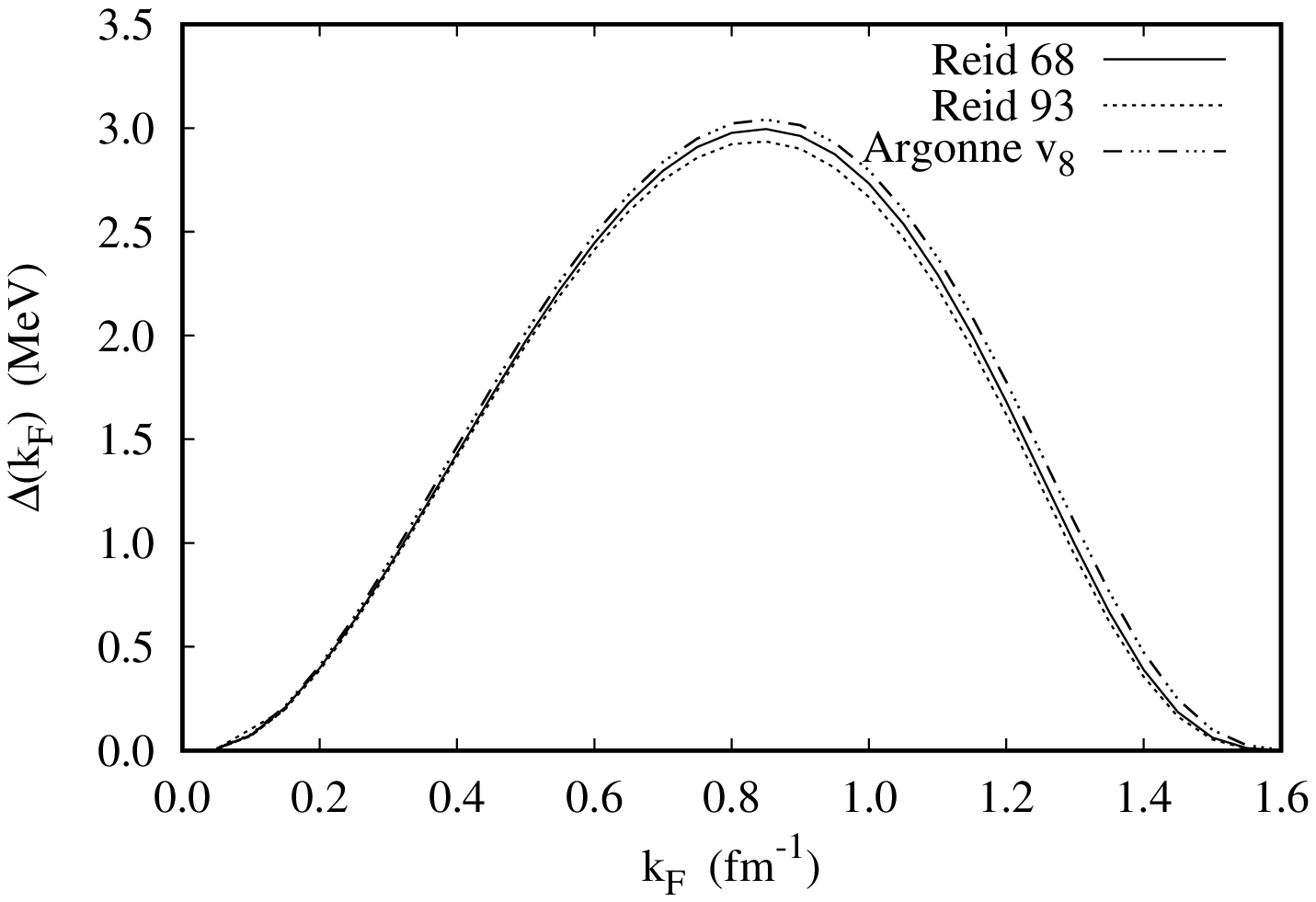}}
  \caption{The figure shows the $^1$S$_0$ gap for the operator form of
    the Reid-68, Reid-93 and the Argonne $V_8$ potentials. The two
    operator versions of Reid-93 give the same
    result.\label{fig:baregaps1S0}}
\end{figure}
\begin{figure}
  \centerline{\includegraphics[width=0.6\columnwidth,angle=270]{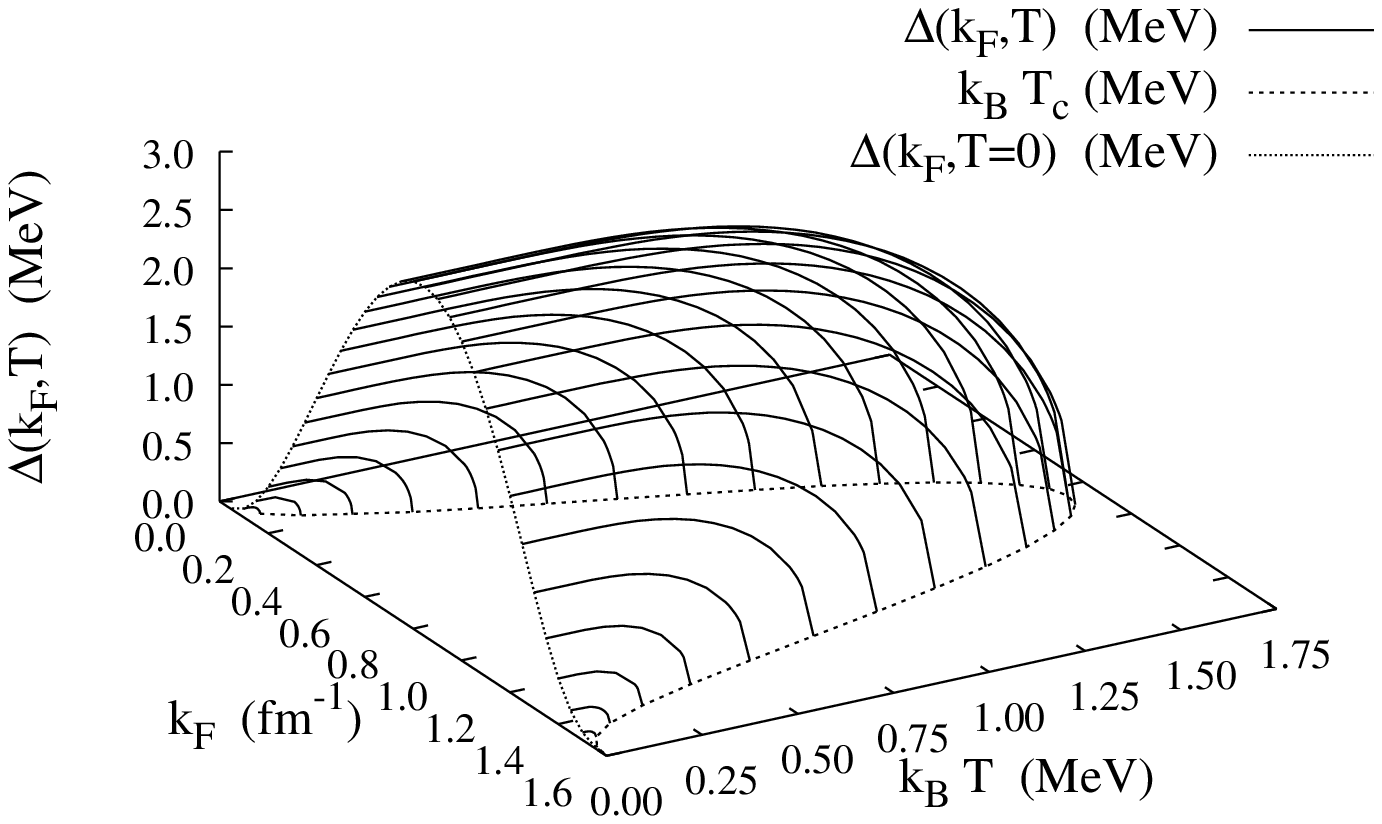}}
\caption{The figure shows, for the Reid 68 interaction, the
  temperature dependence of the $^1$S$_0$ gap. Also shown are the
  critical temperature $T_c$ as a function of density (dashed line) in
  the $(\KF,\KB T)$ plane and the zero temperature solution (dotted
    line in the $(\KF,\Delta(\KF))$ plane).  \label{fig:re68gap1S0T}}
\end{figure}
\begin{figure}
  \centerline{\includegraphics[width=0.6\columnwidth,angle=270]{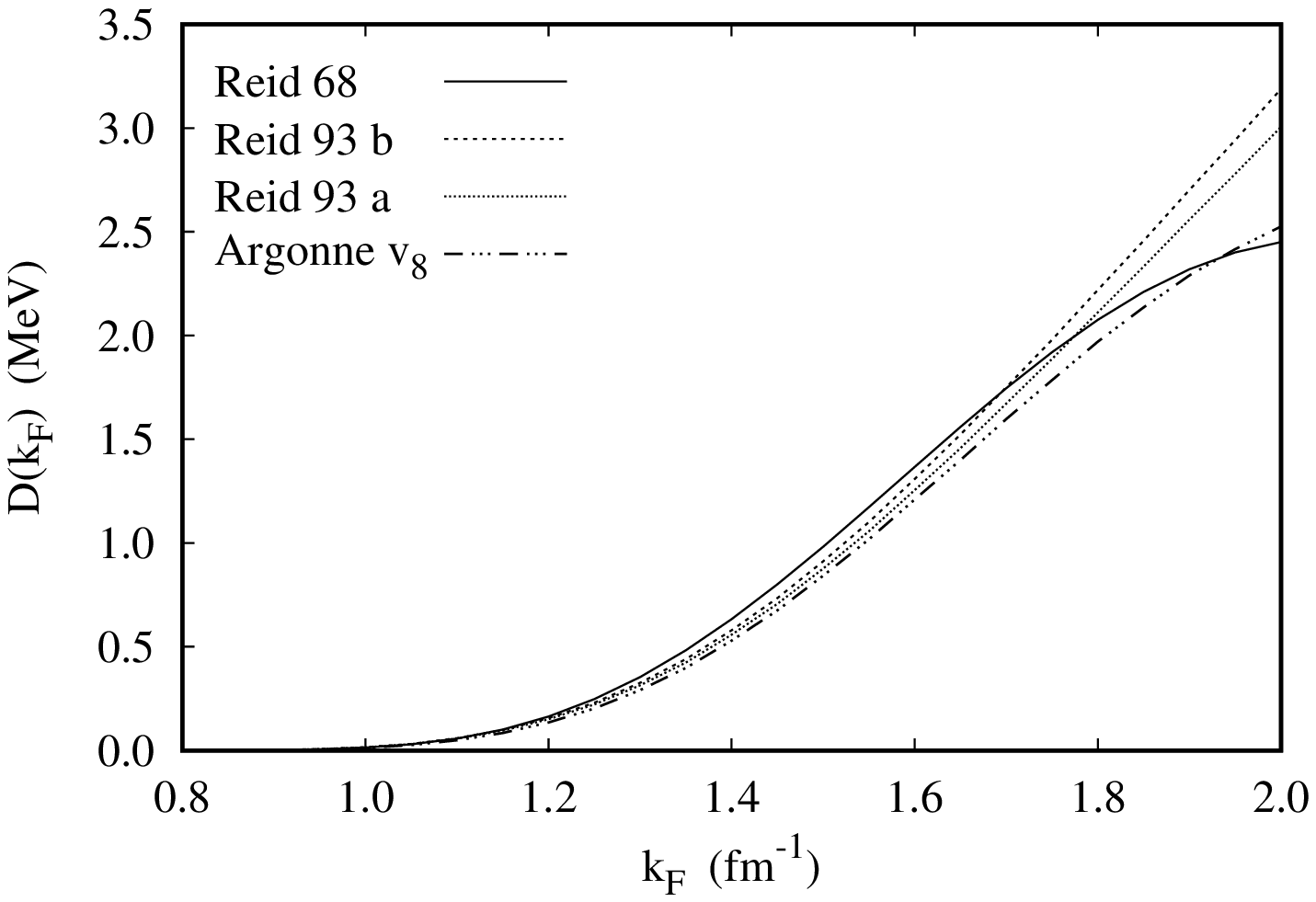}}
\caption{ The figure shows the superfluid coupled channel
  $^3$P$_2$-$^3$F$_2$ gap
  for the four interactions considered here. $D(\KF)$
  is the angle-averaged gap function in the denominator of
  Eq. \eqref{eq:multigap}.\label{fig:baregap3P2-3F2}}
\end{figure}
\begin{figure}
  \centerline{\includegraphics[width=0.6\columnwidth,angle=270]%
  {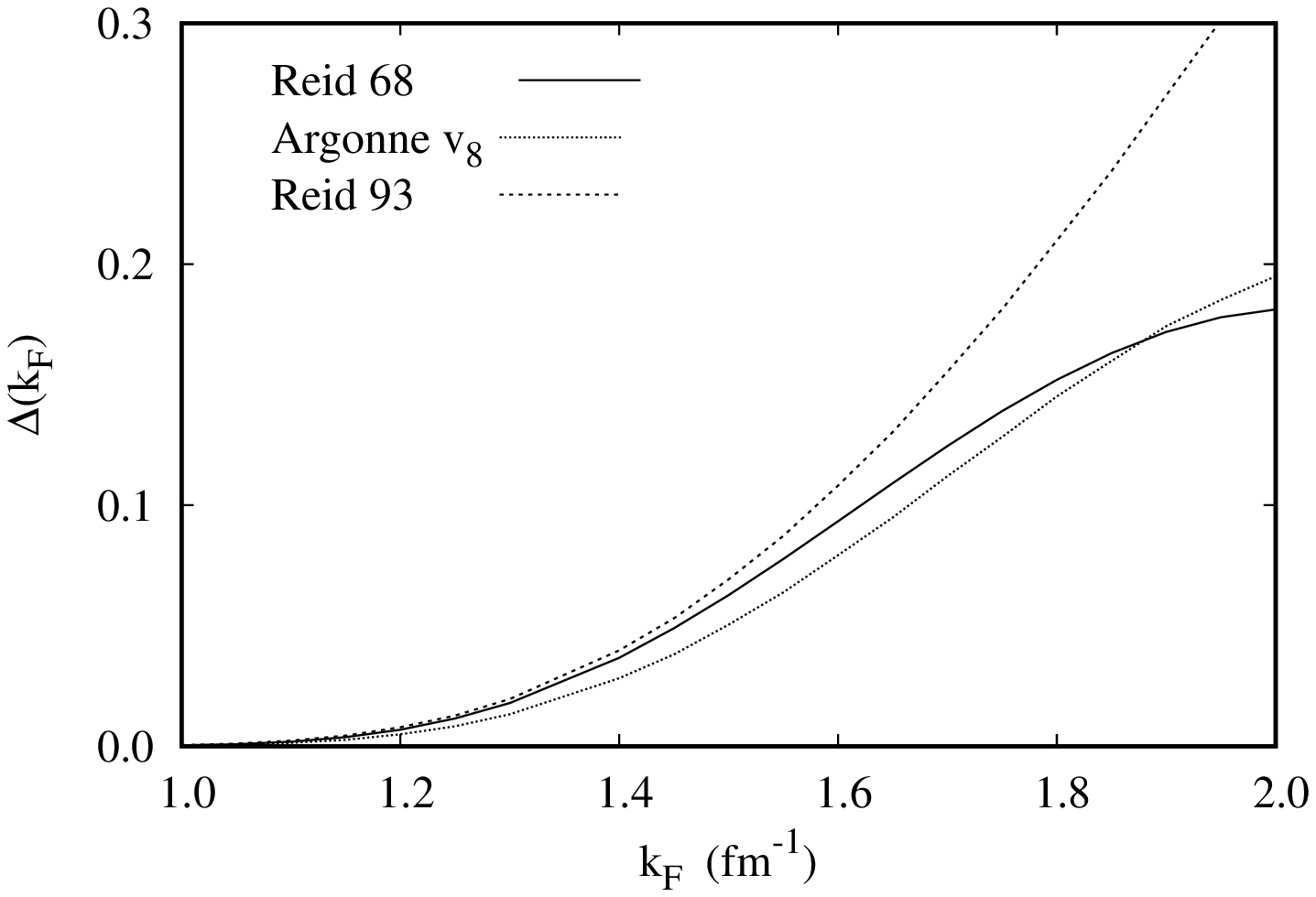}}
\caption{The figure shows the superfluid gap for $^3$P$_2$
  single channel. The two versions of the Reid 93 interaction give
  by construction the same answer. \label{fig:baregap3P2}}
\end{figure}
\begin{figure}
  \centerline{\includegraphics[width=0.6\columnwidth,angle=270]{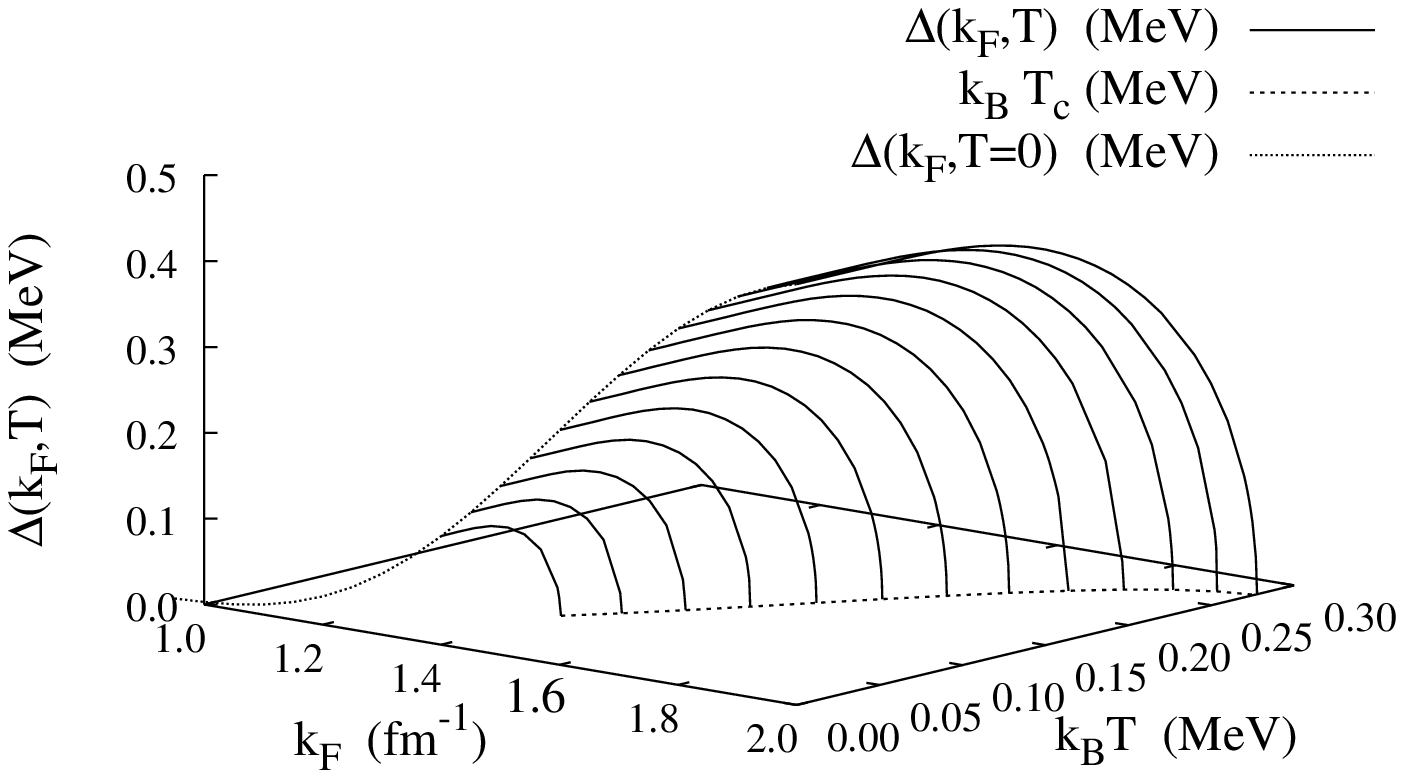}}
  \caption{Same as Fig. \ref{fig:re68gap1S0T} for $^3$P$_2$- $^3$F$_2$
    pairing. \label{fig:re68gap3P2-3F2T}}
\end{figure}
\begin{figure}
  \centerline{\includegraphics[width=0.6\columnwidth,angle=270]{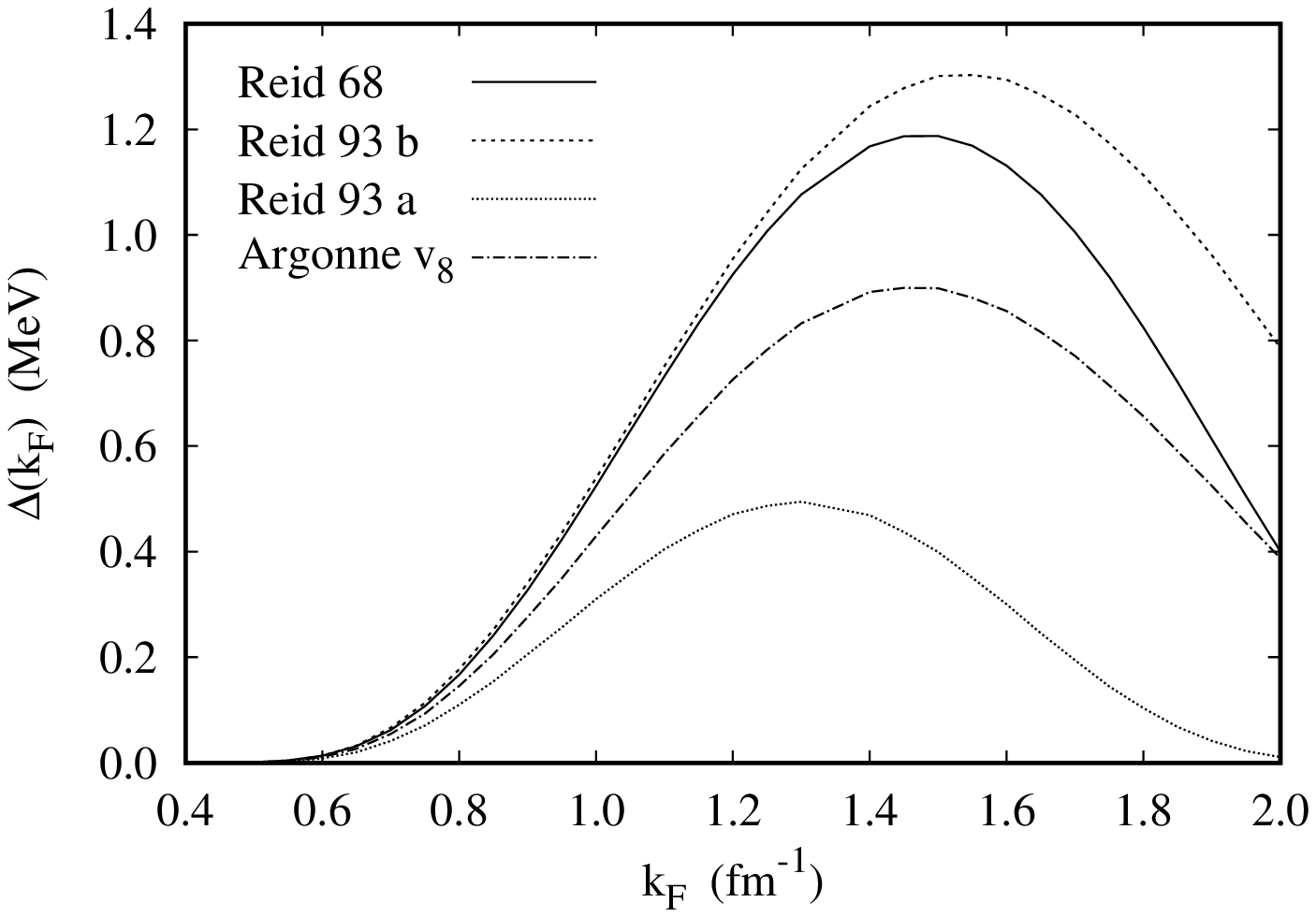}}
  \caption{The figure shows the pairing gap in $^3$P$_0$ states for the
    four interactions considered here when the spin-orbit force has
    been turned off.
 \label{fig:baregaps3P0_noLS}}
\end{figure}

\clearpage 
\mbox{~} 

\bibliography{papers}
\bibliographystyle{apsrev4-2}

\end{document}